\documentclass[%
 aip,
 amsmath,amssymb,
preprint,
]{revtex4-1}
\usepackage{xcolor}
\usepackage{graphicx}
\usepackage{dcolumn}
\usepackage{bm}

\usepackage[utf8]{inputenc}
\usepackage[T1]{fontenc}
\usepackage{mathptmx}
\usepackage{multirow}
\usepackage{etoolbox}

\makeatletter
\def\@email#1#2{%
 \endgroup
 \patchcmd{\titleblock@produce}
  {\frontmatter@RRAPformat}
  {\frontmatter@RRAPformat{\produce@RRAP{*#1\href{mailto:#2}{#2}}}\frontmatter@RRAPformat}
  {}{}
}%
\makeatother
\begin{document}

\title{Nonlinear Reduced-Order Modeling of Compressible Flow Fields Using Deep Learning and Manifold Learning}
\author{Bilal Mufti}
\affiliation{Daniel Guggenheim School of Aerospace Engineering, Georgia Institute of Technology,  Atlanta, Georgia, 30332, USA}
\author{Christian Perron}
\affiliation{Daniel Guggenheim School of Aerospace Engineering, Georgia Institute of Technology, Atlanta, Georgia, 30332, USA}

\author{Dimitri N. Mavris}
\affiliation{Daniel Guggenheim School of Aerospace Engineering, Georgia Institute of Technology, Atlanta, Georgia, 30332, USA}
\date{\today}

\begin{abstract}

This paper presents a nonlinear reduced-order modeling (ROM) framework that leverages deep learning and manifold learning to predict compressible flow fields with complex nonlinear features, including shock waves. The proposed DeepManifold (DM)-ROM methodology is computationally efficient, avoids pixelation or interpolation of flow field data, and is adaptable to various grids and geometries. The framework consists of four main steps: First, a convolutional neural network (CNN)-based parameterization network extracts nonlinear shape modes directly from aerodynamic geometries. Next, manifold learning is applied to reduce the dimensionality of the high-fidelity output flow fields. A multilayer perceptron (MLP)-based regression network is then trained to map the nonlinear input and output modes. Finally, a back-mapping process reconstructs the full flow field from the predicted low-dimensional output modes. DM-ROM is rigorously tested on a transonic RAE2822 airfoil test case, which includes shock waves of varying strengths and locations. Metrics are introduced to quantify the model's accuracy in predicting shock wave strength and location. The results demonstrate that DM-ROM achieves a field prediction error of approximately 3.5\% and significantly outperforms reference ROM techniques, such as POD-ROM and ISOMAP-ROM, across various training sample sizes. 
\end{abstract}


\maketitle

\section{Introduction}\label{sct: intro}
Future aviation designs are expected to operate efficiently at transonic and supersonic speeds while adhering to stringent specific environmental constraints, such as reduced emissions and lower noise levels~\cite{NASA2023,IATA2020}. To meet these challenging demands, aircraft designers are compelled to explore revolutionary designs and configurations. The development of these configurations requires the use of high-fidelity numerical simulations early in the conceptual design phase. For aerodynamic design, computational fluid dynamics (CFD) analysis is commonly employed to evaluate high-speed compressible flow fields. These physics-rich flow fields are essential for determining aerodynamic performance~\cite{toor2020comparative}, conducting multi-disciplinary analysis and optimization~\cite{kenway2014multipoint,Boncoraglio2021ActiveOptimization}, estimating noise~\cite{de2014global}, studying stall characteristics~\cite{khan2019analysis}, and integrating propulsion systems with the airframe~\cite{mufti2019flow,mufti2023selection}. In transonic and supersonic designs, flow fields are characterized by shock waves of varying strength and position, as well as interactions between shock waves and boundary layers. These factors significantly influence aircraft system-level metrics. However, using CFD simulations during the conceptual design phase incurs substantial computational costs and impractical execution times, especially in a many-query context that requires numerous evaluations of these simulations~\cite{martins2021engineering,Forrester2009}. This underscores the need for methods that can predict these flow fields rapidly and at a reduced computational cost for use in aerodynamic design.

Data-driven, non-intrusive reduced-order models (ROMs) offer a promising alternative to high-fidelity CFD simulations~\cite{brunton2020machine}. Unlike intrusive ROMs, which interface with the underlying governing equations of high-fidelity simulations and typically require access to the simulation source code, non-intrusive ROMs are trained using only input and output data, treating the simulation code as a black box~\cite{Lucia2004,Xiao2015Non-intrusiveEquations}. ROMs work by finding a low-dimensional representation, called the \emph{latent space}, of the original high-dimensional flow field. The latent space is constructed to capture the dominant behavior and trends in the original high-dimensional flow field. Once trained, these ROMs can predict flow fields in a computationally efficient manner. In aerodynamic design analysis, the underlying geometry often requires parameterization to enable effective analysis and optimization. Common techniques for this parameterization include splines~\cite{Masters2016}, class shape transformation (CST)~\cite{Kulfan2008}, Hicks-Henne bump functions~\cite{Hicks1978}, and free-form deformation (FFD)~\cite{Kenway2010}. In the context of aerodynamic design, a ROM can generally be considered a predictive map that takes shape parameterization variables and flow conditions as inputs to predict the output flow field~\cite{Perron2021Multi-fidelityAlignment}.

Proper orthogonal decomposition (POD) is a common method used in most ROM techniques, aiming to find a linear latent space of the original high-dimensional space. Due to its simple formulation, minimal computational cost, low training data needs, and a well-defined back mapping, POD has been successfully applied in many aerodynamics and fluid mechanics applications~\cite{Fossati2015EvaluationMethodology,Rajaram2020Non-intrusiveAlgorithms,min2024data,zhang2024prediction}. However, due to its linear nature, POD-ROM struggles to accurately predict flow fields with discontinuities, such as shock waves~\cite{mufti2024multifidelity,Perron2021Multi-fidelityAlignment}. Recently, nonlinear ROMs based on manifold learning have been used to predict field solutions with nonlinearities~\cite{Franz2014Interpolation-basedLearning}. Unlike linear ROMs, these nonlinear techniques attempt to find a lower-dimensional nonlinear manifold that effectively represents the high-dimensional flow field. Decker~\textit{et al.}\cite{Decker2023ManifoldModeling} developed various single- and multi-fidelity nonlinear ROMs using Kriging regressors. They tested these ROMs on transonic and supersonic airfoil test cases parameterized using the FFD technique. The results showed that manifold learning-based nonlinear ROMs performed better than their linear counterparts in the vicinity of shock waves but did not substantially improve the overall field prediction error. Their findings also revealed that as the number of FFD points or the dimension of the input space increases, the resulting ROMs started to suffer from reduced accuracy due to the \textit{curse of dimensionality}~\cite{Mufti2023DesignSubspaces}. Iyengar~\textit{et al.}\cite{iyengar2024domain} used a domain decomposition technique to enhance the performance of manifold learning-based ROMs by developing heuristics to divide the flow domain into different subdomains. This approach allowed the application of linear and nonlinear techniques in different subdomains to construct a single nonlinear predictive ROM. However, they observed areas of large errors around subdomain interfaces. 

Deep learning techniques have recently achieved success in the field of fluid mechanics and aerodynamics by effectively modeling complex systems. They have been used to study unsteady flows around bluff bodies~\cite{Eivazi2020DeepFlows}, active flow control~\cite{liu2024novel}, heat transfer analysis~\cite{cai2021physics}, and flow turbulence modeling~\cite{Nakamura2021ConvolutionalFlow}. Deep learning techniques have also been extensively applied to study aerodynamic flow fields. Techniques utilizing deep neural networks (NNs) have been developed to predict both incompressible and compressible flow fields. These techniques take shape parameters of aerodynamic geometry, flow conditions, and physical coordinates of the flow field as input to the NN and predict the flow field as output~\cite{sekar2019fast,Sun2021ASpeeds}. This point-by-point approach treats each output mesh node as an independent prediction point, which does not take advantage of the overall structure of the flow field. Consequently, this method potentially limits prediction accuracy by overlooking the spatial relationships and patterns between nodes. 

Convolutional neural networks (CNNs) utilize this spatial information through convolution operations. Thuerey~\textit{et al.}~\cite{Thuerey2020DeepFlows} used a CNN-based UNet architecture to predict pressure and velocity fields for different airfoil designs and flow conditions. The UNet architecture takes an image of the airfoil shape as input, with the encoder downsampling it to extract latent nonlinear modes, while the decoder uses upsampling operations to reconstruct the flow field. Different distance functions have also been introduced as additional inputs to CNNs to improve their prediction accuracy~\cite{Bhatnagar2019PredictionNetworks,Duru2021CNNFOIL:Airfoils}. DIP-ShockNet, a domain-informed and probabilistic CNN-based architecture, has recently been introduced to enhance prediction accuracy in regions with shock waves by using flow field gradient information~\cite{mufti2024shock}. However, these studies mapped the original CFD grid to a Cartesian grid to enable the CNN-based architecture to train. This \emph{pixelation} strategy introduces interpolation errors in the flow field, loss of information in the near-wall region, and artificial roughness on smooth boundaries. These issues lead to oscillations in the flow field close to the geometry boundary and smoothing of nonlinearities such as shock waves.

Domain transformation techniques can help alleviate the effects of pixelation when using CNN-based architectures. These techniques transform the non-uniform flow field into a uniform Cartesian computational plane where convolution operations are performed~\cite{Sun2021ASpeeds, Chen2023TowardsAerofoils, Deng2023PredictionStrategies}. However, these mesh transformation techniques require the underlying CFD grid to have a specific topological structure, making them unsuitable for unstructured grids, such as those developed for complex aerodynamic shapes. Graph neural networks (GNNs) allow direct use of unstructured grids by employing graph theory to represent the grid~\cite{Hall2021GINNs:Physics,Pfaff2020LearningNetworks}. However, GNNs face limitations due to high memory requirements and inefficient parallelization for large graphs, making them less practical for high-fidelity aerodynamic applications.~\cite{Deng2023PredictionStrategies,Chen2023TowardsAerofoils}. A point-cloud deep learning technique for predicting flow fields on various geometries has been introduced by Kashefi~\textit{et al.}~\cite{Kashefi2021AGeometries}. The PointNet architecture captures the locations of grid vertices and uses them as inputs to the network. This technique has successfully predicted flow fields in irregular domains but has not been extensively tested for high-speed compressible flows.

To overcome the limitations of existing reduced-order modeling and deep learning techniques in predicting high-speed aerodynamic flow fields, we introduce a novel nonlinear reduced-order modeling framework, which leverages the strengths of deep learning and manifold learning. The proposed framework consists of four key components. First, a CNN-based parameterization network is developed to extract nonlinear shape modes using only geometric shape information. Second, we employ manifold learning, specifically the Isometric Mapping (ISOMAP) technique, to capture a low-dimensional manifold representing the high-dimensional flow field, thereby handling the complex flow variations. Third, a multilayer perception (MLP) regression network is trained to map the extracted nonlinear shape modes and flow conditions to the low-dimensional output space. Finally, a back-mapping process is employed to project the low-dimensional predictions back into the original high-dimensional flow field space.

The primary contributions of our proposed methodology can be summarized as follows:
\begin{itemize}
    \item The proposed methodology offers an efficient, end-to-end nonlinear ROM framework based on deep learning and manifold learning to predict aerodynamic flow fields with nonlinearities, such as shock waves.
    \item The framework does not require pixelation or interpolation of the flow field data, preserving the accuracy of the aerodynamic data during training and prediction.
    \item The methodology is adaptable to any underlying CFD grid, including O-grid, C-grid, and unstructured grids, making it flexible for a wide range of applications.
    \item Unlike other nonlinear ROM techniques, the methodology does not require an a priori parameterization of aerodynamic geometries. The CNN-based parameterization network extracts the underlying nonlinear parameterization from the aerodynamic geometry information. Additionally, the framework is less sensitive to the curse of dimensionality, even as the input space grows.
    \item We introduce metrics to rigorously quantify the accuracy of shock wave predictions, measuring both the strength and location of shock waves in the flow field.
\end{itemize}

The remainder of this paper is organized as follows. Section~\ref{sct: methodology} provides a detailed description of the framework, including the design of the CNN-based parameterization network, the dimensionality reduction using the ISOMAP technique, and the construction of the MLP regression network for predicting the low-dimensional flow field. The back-mapping process to recover the high-dimensional flow field from the latent space is also explained. Section~\ref{sct:application} introduces the transonic airfoil test case used in this study, outlines the data generation process for both geometric shapes and flow fields, and defines the performance metrics developed to evaluate the accuracy of the predictions, especially in shock wave regions. Section~\ref{sct: results} presents the training results of the CNN-based parameterization and MLP regression networks and compares the proposed ROM methodology with reference ROM techniques. Finally, Section~\ref{sct:conclusion} summarizes the key findings and discusses potential future extensions.

\section{Methodology}\label{sct: methodology}
In this section, we present the proposed nonlinear ROM methodology in detail. The framework consists of four key components: the CNN-based shape parameterization network, the ISOMAP manifold learning nonlinear dimensionality reduction technique, the MLP-based regression, and the back-mapping methodology. We will describe the architecture of the deep learning models used and introduce the manifold learning technique. The section concludes by integrating these key elements to develop the proposed nonlinear ROM framework.

\subsection{CNN-based Parameterization Network}\label{subsct:CNN-model}
A CNN is a specialized deep learning model primarily used for computer vision applications. Its architecture significantly reduces the number of parameters in the network, making it efficient for feature extraction from large-scale structured inputs~\cite{lecun2010convolutional}. CNNs have been utilized to predict aerodynamic coefficients~\cite{zhang2018application,yu2019improved} and for the inverse design of airfoils~\cite{sekar2019inverse}. In this study, we employ a CNN architecture to derive shape modes from a set of training samples. These modes differ from those obtained through conventional shape parameterization methods discussed in Section~\ref{sct: intro}, where the design variables locally deform the aerodynamic shape. From a given aerodynamic shape input, the CNN-based parameterization network reduces the details of the aerodynamic geometry to a few useful nonlinear modes~\cite{zuo2022fast}. These nonlinear shape modes potentially offer a more efficient parameterization compared to conventional methods~\cite{li2022machine}. The CNN-based parameterization approach is easy to implement, computationally efficient, and generalizable to various shapes.

A CNN architecture typically consists of convolutional layers, pooling layers, and fully connected layers. The convolution layers use filters or convolutional kernels to perform convolutions on the input data, producing a feature map. The nonlinear activation function helps in capturing more complex relationships and patterns in the input. Mathematically, the convolution process can be expressed as:

\begin{equation}
\mathbf{C}_{j} = \sigma(\mathbf{f}_j \ast \mathbf{x} + b_j), \quad j = 1, 2, \ldots, L,
\end{equation}
where \(\mathbf{x}\) represents the input tensor to the convolutional layer with dimensions \(W_{i} \times H_{i}\). The symbol \(\ast\) denotes the convolution operation, \(\mathbf{f}_j\) is the \(j\)-th convolutional filter with size \(F \times F\), and \(b_j\) is the bias associated with the \(j\)-th filter. The function \(\sigma\) denotes a nonlinear activation function. Common choices for activation functions include ReLU, sigmoid, or tanh. The output matrix corresponding to the \(j\)-th convolutional operation is denoted by \(\mathbf{C}_{j}\). The number of filters in the layer is given by \(L\). The width \(W_{o}\), and height \(H_{o}\) of the output tensor are determined by the size of the filters and the stride used in the convolution operation, which is typically set to \(s=1\). For each filter, the output matrix \(\mathbf{C}_{j}\) is obtained by taking the dot product of the corresponding sub-region of the input tensor \(\mathbf{x}\) and the convolutional filter \(\mathbf{f}_{j}\). Pooling layers downsample the dimension of the data using a pooling operation, which helps reduce the number of parameters and computations in the network and avoids overfitting~\cite{xiao2021addressing}. Max pooling is one of the most commonly used pooling operations, which takes the maximum value in a pooling window while reducing the spatial size of the input information. The fully connected layers in the network are dense layers of neurons followed by activation functions.

A representation of the CNN-based parameterization network used in this study is shown in Fig~\ref{fig:cnn_network}. The network uses an encoder similar to the one used in autoencoders, but instead of recovering the aerodynamic shape image as output, it predicts the coordinates of the geometry. The network's encoder part uses convolutional and fully connected layers to extract the nonlinear shape parameters. The number of shape parameters \(p\) is controlled by the size of the parameter layer. A fully connected network decodes these shape parameters into coordinates corresponding to geometric shapes. The model is trained using the mean squared error (MSE) loss function, which is defined as:

\begin{figure*}[]
    \centering
    \includegraphics[width=1\textwidth]{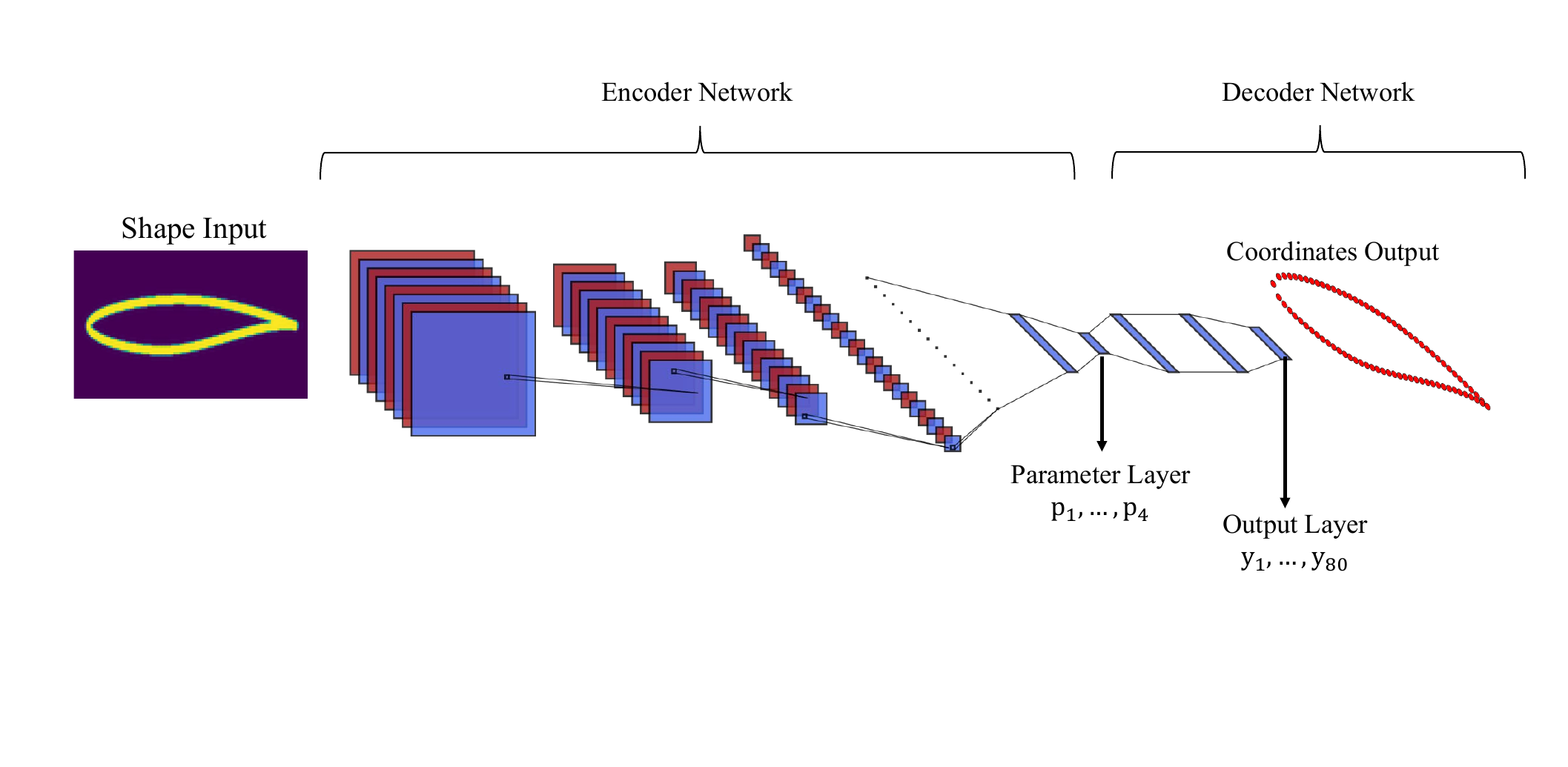}
    \caption{A generic representation of CNN-based parameterization network.}
    \label{fig:cnn_network}
\end{figure*}

\begin{equation}\label{eq:cnn_rms}
\text{loss}_{\text{CNN}} = \frac{1}{n_{s} \cdot n_{c}} \sum_{i = 1}^{n_{s}}\sum_{j = 1}^{n_{c}}  (y_{i,j} - \tilde{y}_{i,j})^2
\end{equation}
where \(y_{i,j}\) and \(\tilde{y}_{i,j}\) represent the actual and predicted \(y\)-coordinates, respectively, \(n_{s}\) is the number of samples, and \(n_{c}\) is the number of \(y\)-coordinates in each sample.

We developed and compared three CNN-based parameterization networks to identify the most suitable architecture for nonlinear feature extraction. Table~\ref{tab:architecture_selection} summarizes the architectures of these networks, labeled as A1, A2, and A3, with increasing complexity. The A1 network consists of four convolutional layers with a varying number of filters and nine fully connected layers. The number of neurons in the parameterization layer corresponds to the number of shape parameters \(p\) to be extracted, while the output layer's neurons match the number of coordinates of the aerodynamic geometry. The A2 network increases the complexity by adding more convolutional layers while maintaining the same number of fully connected layers as A1 but with a higher number of neurons in each layer. The A3 network further increases the complexity, incorporating additional convolutional filters in each convolutional layer and increasing the number of neurons in the fully connected layers.

Each convolutional layer in the encoder uses a 4 $\times$ 4 convolutional filter with a stride of 1, chosen to capture sufficient spatial features without excessive computational overhead. The ReLU activation function is applied in all convolutional layers due to its efficiency and effectiveness in mitigating the vanishing gradient problem. To reduce input dimensionality and prevent overfitting, 2D max pooling is applied in the last four convolutional layers of each network. The fully connected layers employ the tanh activation function, which effectively bounds the output between -1 and +1, aligning with the scaled coordinates of the aerodynamic geometry. Additionally, kernel regularization with an L2 penalty is implemented in the fully connected layers to further reduce overfitting.

\begin{table*}[ht]
\caption{\label{tab:architecture_selection} Summary of different architectures for CNN-based parameterization networks. ($p$ = Number of shape parameters, $O$ = Number of output coordinates)}
\begin{ruledtabular}
\begin{tabular}{@{}lcccc@{}}

Architecture & \multicolumn{2}{c}{Convolutional layers} & \multicolumn{2}{c}{Fully connected layers}                   \\ 
             & No. of layers & No. of filters         & No. of layers & No. of neurons in each layer                 \\ 
\hline
A1           & 4             & [64, 64, 128, 256]         & 9             & 256, 2 $\times$ 128, $p$, 4 $\times$ 100, $O$              \\
A2           & 6             & [32, 64, 64, 128, 256, 512]   & 9             & 512, 2 $\times$ 200, $p$, 4 $\times$ 300, $O$              \\
A3           & 6             & [64, 64, 128, 256, 512, 1024] & 12       & 1024, 512, 3 $\times$ 200, $p$, 5 $\times$ 500, $O$ \\ 
\end{tabular}
\end{ruledtabular}
\end{table*}


\subsection{Manifold Learning using ISOMAP}\label{subsct:manifold_learning}
Manifold learning algorithms are a subset of nonlinear dimensionality reduction techniques~\cite{van2009dimensionality}. These methods are designed to uncover a lower-dimensional nonlinear manifold embedded within a higher-dimensional data space. The primary objective of manifold learning is to identify and capture the underlying structure of high-dimensional data by reducing its dimensionality while preserving essential geometric properties. These methods can broadly be categorized into global and local techniques~\cite{izenman2012introduction}. Global techniques aim to maintain relationships across the entire dataset, while local methods focus on preserving local neighborhood structures. Among the various methods developed, one prominent global approach is Isometric Mapping (ISOMAP). This technique has demonstrated superior prediction accuracy for ROMs developed for nonlinear systems compared to local techniques such as Local Tangent Space Alignment (LTSA)~\cite{Decker2023ManifoldModeling}.

ISOMAP extends the classical Multidimensional Scaling (MDS) by focusing on preserving the geodesic distances between all pairs of data points. Geodesic distances represent the shortest path between points on a curved manifold, providing a more accurate measure of similarity for nonlinearly structured data. The following sections outline the key steps of the ISOMAP technique from the formulation provided by Franz~\textit{et al.}~\cite{Franz2014Interpolation-basedLearning}.

\subsubsection{Constructing the neighborhood graph}

The first step in the ISOMAP algorithm involves representing the dataset as a weighted graph, which serves as a discrete approximation of the underlying manifold. Given a dataset $\mathbf{Y} = \{\mathbf{y}_1, \mathbf{y}_2, \dots, \mathbf{y}_n\} \subset \mathbb{R}^{m}$, each data point is treated as a vertex in the graph. Edges between these vertices are established based on proximity in the high-dimensional space, which can be determined using either the $k$-nearest neighbors (kNN) or the $\epsilon$-neighborhood criterion.

In the $k$-nearest neighbors approach, each point is connected to its $k$ nearest neighbors, ensuring a consistent number of edges per vertex. Alternatively, in the $\epsilon$-neighborhood approach, each point is connected to all other points within a radius $\epsilon$. The choice of $k$ or $\epsilon$ significantly influences the graph's structure and, consequently, the accuracy of the manifold's representation. The edges are weighted according to the Euclidean distances between the connected points, represented as $d_{ij} = \|\mathbf{y}_i - \mathbf{y}_j\|$. This graph representation captures the neighborhood relationships in the dataset, approximating the manifold's structure by assuming that locally, the Euclidean distance is a good proxy for the geodesic distance on the manifold.

In this study, we adopt the kNN approach due to its straightforward implementation and the control it offers over the neighborhood size. However, selecting the optimal number of neighbors $k$ is a critical aspect, as it is a hyperparameter that can significantly influence the results. An inappropriate choice of $k$ can either lead to disconnected components or overly dense connections, distorting the manifold's representation. To address this, we utilize the method proposed by Decker~\textit{et al.}~\cite{Decker2023ManifoldModeling}, which employs Kruskal stress minimization to determine the optimal number of neighbors $k^*$. This optimal value of $k^*$ is selected to preserve the relative geodesic distances after projecting the data into the latent space, ensuring that the manifold's intrinsic geometry is accurately represented.

\subsubsection{Computing geodesic distances}

Following the construction of the neighborhood graph, the ISOMAP algorithm computes the geodesic distances between all pairs of data points. Geodesic distances represent the shortest paths along the manifold and are crucial for preserving the manifold's true geometric properties. In the graph representation, these distances are approximated by the shortest path distances between vertices. The computation of these distances involves using shortest-path algorithms, such as Dijkstra's~\cite{dijkstra2022note} or Floyd-Warshall's~\cite{floyd1962algorithm} algorithms, which calculate the minimum cumulative distance along paths in the graph. Specifically, the geodesic distance $d_G(\mathbf{y}_i, \mathbf{y}_j)$ between points $\mathbf{y}_i$ and $\mathbf{y}_j$ is determined by minimizing the sum of the edge weights along the path $P_{ij}$ that connects these points:

\begin{equation}
d_G(\mathbf{y}_i, \mathbf{y}_j) = \min_{P_{ij}} \sum_{(p,q) \in P_{ij}} d_{pq}
\end{equation}

The resulting distances are stored in a geodesic distance matrix \(D_G\), which represents the pairwise geodesic distances for all points in the dataset. 

\subsubsection{Creating the low-dimensional embedding}

The final step in the ISOMAP methodology involves creating a low-dimensional embedding that preserves the computed geodesic distances. This process begins with the transformation of the geodesic distance matrix \(D_G\) into a matrix of inner products through double-centering, achieved using the equation:

\begin{equation}
B = -\frac{1}{2} J D_G^2 J
\end{equation}
where \(J = I - \frac{1}{n} \mathbf{1}\mathbf{1}^T\) is the centering matrix, \(I\) is the identity matrix, and \(\mathbf{1}\) is a vector of ones. The resulting matrix \(B\) is then subjected to eigenvalue decomposition, yielding a set of eigenvalues \(\{\lambda_1, \lambda_2, \ldots, \lambda_m\}\) and corresponding eigenvectors \(\{\mathbf{v}_{1}, \mathbf{v}_{2}, \ldots, \mathbf{v}_{m}\}\). The largest \(d\) eigenvalues (where \(d \ll m\)) and their associated eigenvectors are selected to form the basis of the low-dimensional space. The coordinates of the data points in this new low-dimensional space, \(\mathbf{Z} = \{\mathbf{z}_1, \mathbf{z}_2, \dots, \mathbf{z}_n\} \subset \mathbb{R}^{d}\), are calculated using these eigenvectors, providing a representation that preserves the geodesic distances as closely as possible.


\subsection{MLP-based Regression Network}

The output of the ISOMAP technique is a projection matrix \(\mathbf{Z}\) containing coordinates of high-dimensional field data projected onto a low-dimensional latent space \(d\). The next step involves developing a parametric map \(h:\mathbb{R}^{p} \mapsto \mathbb{R}^{d}\) to predict the latent space for different input samples. Various regression techniques, including linear regression~\cite{montgomery2021introduction}, radial basis functions (RBFs)~\cite{Han2012HierarchicalModeling}, and Kriging~\cite{Fossati2015EvaluationMethodology}, have been used previously.

\begin{figure*}[!ht]
    \centering
    \includegraphics[width=0.75\textwidth]{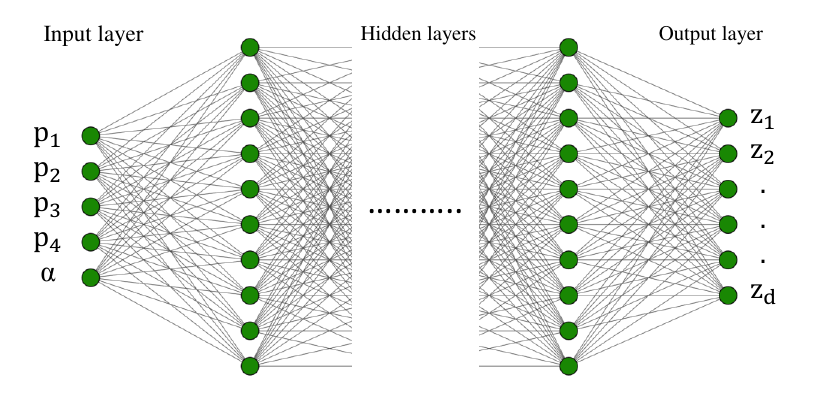}
    \caption{A generic representation of MLP-based regression network.}
    \label{fig:mlp_network}
\end{figure*}

In this study, we employ a Multilayer Perceptron (MLP) neural network. An MLP consists of an input layer, one or more hidden layers, and an output layer as shown in Fig.~\ref{fig:mlp_network}. The input layer receives the input features, which are transformed through linear combinations and activation functions across multiple layers. The output layer produces the final prediction. Mathematically, the operation of an MLP can be expressed as:

\begin{equation}
\mathbf{z}^{(l)} = \sigma^{(l)}(\mathbf{W}^{(l)} \mathbf{z}^{(l-1)} + \mathbf{b}^{(l)}), \quad l = 1, 2, \ldots, L,
\end{equation}
where \(\mathbf{z}^{(0)}\) represents the input vector of features, with \(\mathbf{z}^{(0)} \in \mathbb{R}^{p}\). The weight matrix for the \(l\)-th layer, \(\mathbf{W}^{(l)}\), linearly transforms the input from the previous layer, with dimensions determined by the number of neurons in the \(l-1\)-th and \(l\)-th layers. The bias vector \(\mathbf{b}^{(l)}\) allows the model to better fit the data by shifting the activation function. The nonlinear activation function \(\sigma^{(l)}\) is applied element-wise to the result of the linear transformation in the \(l\)-th layer. Finally, \(\mathbf{z}^{(l)}\) represents the output vector of the \(l\)-th layer, serving as the input to the subsequent layer. This process continues through the network layers until the final layer (\(L\)), where the output vector \(\mathbf{z}^{(L)}\) represents the prediction in the latent space. The transformations and activation operations at each layer enable the MLP to learn complex, nonlinear mappings from the input to the output space, making it a powerful tool for regression tasks.

To determine the most suitable MLP architecture for predicting the output latent space coordinates, we developed and tested seven different MLP regression networks. These networks, summarized in Table~\ref{tab:mlp_selection}, vary in depth and width, achieved by altering the number of fully connected layers and the number of neurons per layer. The architectures range from smaller, simpler networks (e.g., M1 with 10 layers of 100 neurons each) to larger, more complex configurations (e.g., M7 with 10 layers of 1000 neurons each). Each layer employs the ReLU activation function, chosen for its computational efficiency and effectiveness in handling nonlinearities. To mitigate overfitting, L2 kernel regularization is applied across all layers.

\begin{table}[]
\caption{\label{tab:mlp_selection} Summary of different architectures for MLP regression networks.}
\begin{ruledtabular}
\begin{tabular}{@{}lcc@{}}
Architecture &  Fully connected layers & Total neurons \\ 
\hline
M1 & 10 $\times$ 100 & 1,000 \\
M2 & 10 $\times$ 200 & 2,000 \\
M3 & 5 $\times$ 500 & 2,500 \\
M4 & 8 $\times$ 400 & 3,200 \\
M5 & 6 $\times$ 600 & 3,600 \\
M6 & 8 $\times$ 800 & 6,400 \\
M7 & 10 $\times$ 1000 & 10,000 \\
\end{tabular}
\end{ruledtabular}
\end{table}

\subsection{Back-Mapping from Latent Space to High-Dimensional Output}

The final component of our nonlinear ROM framework involves back-mapping the low-dimensional latent space coordinates, obtained through the ISOMAP technique, back to the high-dimensional output space. This process is crucial as we aim to predict the high-dimensional flow field as our output. After obtaining the latent space coordinates \(\mathbf{z}^* \in \mathbb{R}^{d}\) for a new data point, the corresponding high-dimensional output \(\mathbf{y}^* \in \mathbb{R}^{m}\) can be reconstructed. This is achieved using back-mapping methods, which effectively approximate the high-dimensional data from the reduced-order model outputs~\cite{saul2003think}. The approach involves determining a set of weights \(w_j\) by solving an optimization problem, which then linearly combines the high-dimensional snapshots corresponding to the nearest neighbors in the latent space.

Given the \(k\)-nearest neighbors \(\mathbf{z}_j \in \mathbb{R}^{d}\) of the point \(\mathbf{z}^*\) in the latent space, where \(j = 1, 2, \ldots, k\), and their corresponding high-dimensional representations \(\mathbf{y}_j \in \mathbb{R}^{m}\), the reconstruction of \(\mathbf{y}^*\) is formulated as:

\begin{equation}
\mathbf{y}^* = \sum_{j=1}^{k} w_j \mathbf{y}_j,
\end{equation}

The weights \(w_j\) are determined by minimizing the following objective function~\cite{Franz2014Interpolation-basedLearning}, which includes a penalty term to regularize the solution:

\begin{equation}
\min_{\mathbf{w}} \left\|\mathbf{z}^* - \sum_{j=1}^{k} w_j \mathbf{z}_j\right\|^2 + \epsilon \sum_{j=1}^{k} c_j w_j^2, \quad \text{s.t.} \sum_{j=1}^{k} w_j = 1,
\end{equation}
where

\begin{equation}
c_j = \left( \frac{\|\mathbf{z}^* - \mathbf{z}_j\|_2}{\max_i \|\mathbf{z}^* - \mathbf{z}_i\|_2} \right)^\gamma
\end{equation}
and \(\epsilon\) and \(\gamma\) are hyperparameters chosen such that \(0 < \epsilon \ll 1\) and \(1 < \gamma \in \mathbb{N}\). The quantity \(c_j\) is a penalty term applied to the \(j\)-th snapshot that regularizes the optimization and decreases the contribution of snapshots far from \(\mathbf{z}^*\). The solution to this optimization problem is unique and provides the weight coefficients \(w_j^*\) used to reconstruct the high-dimensional data as a linear combination of the corresponding high-dimensional snapshots.

This method assumes that the local neighborhood in the latent space approximately corresponds to the local neighborhood in the high-dimensional space, preserving local structures and ensuring meaningful back-mapping. By carefully selecting the parameters \(\epsilon\), and \(\gamma\), this approach ensures robust and accurate reconstruction of high-dimensional data. We use the optimal values of \(\epsilon = 0.1\), and \(\gamma =4\) outlined in ~\cite{Franz2014Interpolation-basedLearning} for predicting flow fields.

\subsection{Proposed Nonlinear ROM Methodology}

The primary contribution of this work is the development of a computationally efficient, nonlinear reduced-order modeling technique for predicting aerodynamic flow fields with nonlinearities. The methodology integrates deep learning and manifold learning techniques to construct a fast, end-to-end prediction framework. The proposed method, henceforth referred to as DeepManifold-ROM (DM-ROM), is summarized below:

\begin{enumerate}
    \item \textbf{Generate data:} Sample aerodynamic shapes from the input space \(\mathbb{P}\). Run high-fidelity CFD simulations to generate a dataset of aerodynamic flow fields \(\mathbf{Y} = \{\mathbf{y}_1, \mathbf{y}_2, \dots, \mathbf{y}_n\} \subset \mathbb{R}^{m}\).

    \item \textbf{Extract nonlinear shape modes:} Develop a CNN-based parameterization network to extract \(p\) nonlinear shape modes from the aerodynamic shape information. The CNN reduces the aerodynamic geometry details to a few useful nonlinear modes, creating a low-dimensional input space.

    \item \textbf{Identify low-dimensional manifold:} Apply ISOMAP to the flow field data to identify a low-dimensional manifold representing the field solution:
    \begin{enumerate}
        \item Construct a neighborhood graph of the flow field dataset using the \(k\)-nearest neighbors (kNN) approach.
        \item Compute geodesic distances between all pairs of data points on this graph using shortest-path algorithms and store them in a matrix \(D_G\).
        \item Create a low-dimensional embedding \(\mathbf{Z} = \{\mathbf{z}_1, \mathbf{z}_2, \dots, \mathbf{z}_n\} \subset \mathbb{R}^{d}\) that preserves the geodesic distances by performing eigenvalue decomposition on the geodesic distance matrix \(D_G\).
    \end{enumerate}

    \item \textbf{Train regression model:} Develop a parametric map \(h:\mathbb{R}^{p} \mapsto \mathbb{R}^{d}\) using an MLP neural network to predict the latent space coordinates \(\mathbf{z}^*\) for a new input sample.

    \item \textbf{Back-mapping to high-dimensional output:} Reconstruct the high-dimensional output \(\mathbf{y}^*\) from the latent space coordinates \(\mathbf{z}^*\) using a back-mapping technique:
    \begin{enumerate}
        \item Identify the \(k\)-nearest neighbors of \(\mathbf{z}^*\) in the latent space and their corresponding high-dimensional representations.
        \item Solve the optimization problem to determine the weight coefficients \(w_{j}\).
        \item Use the weight coefficients to reconstruct the high-dimensional output as a linear combination of the nearest neighbors using \(\mathbf{y}^* = \sum_{j=1}^{k} w_j \mathbf{y}_j\).
    \end{enumerate}
\end{enumerate}

\section{Application Problem}\label{sct:application}
To evaluate the effectiveness of our proposed framework in predicting flow fields with nonlinearities, such as shock waves, we selected the RAE2822 transonic airfoil as the test case. This airfoil is representative of designs used in high-speed aircraft wings and serves as the baseline geometry for the second benchmark problem provided by the AIAA Aerodynamic Design Optimization Discussion Group (ADODG)~\cite{Lee2015}.

\subsection{Geometric Shape Data Generation}\label{subsct:shape_data_generation}

We generated a diverse set of airfoil designs to train our CNN-parameterization networks using the FFD geometric parameterization technique applied to the baseline RAE2822 airfoil. The FFD technique encloses the aerodynamic geometry in an FFD box, with specific control points defined as nodes on this box. Linear displacement applied to these FFD nodes modifies the shape of the enclosed geometry. It is important to highlight that while the geometric parameterization technique is used here to generate various airfoil designs and to define design variables (FFD nodes) for training reference ROM techniques, the CNN-parameterization network used in our DM-ROM framework does not require an underlying geometric parameterization to extract shape modes. This flexibility allows the use of databases like the UIUC database, containing various airfoils without a unique underlying parameterization, to train the CNN-parameterization network within the DM-ROM methodology.

\begin{figure*}[ht]
    \centering
    \includegraphics[width=1\textwidth]{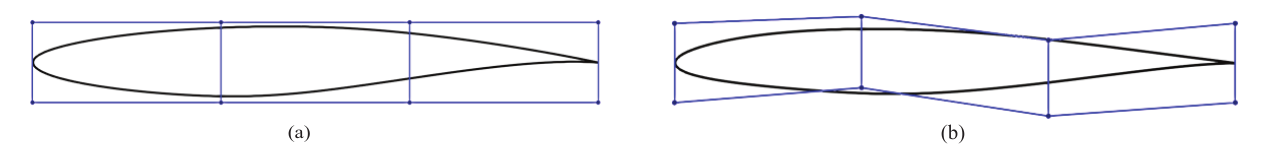}
    \caption{Shape parametrization of RAE2822 using FFD technique: (a) original geometry and, (b) deformed geometry.}
    \label{fig:FFD}
\end{figure*}

\begin{figure}[]
    \centering
    \includegraphics[width=0.65\textwidth]{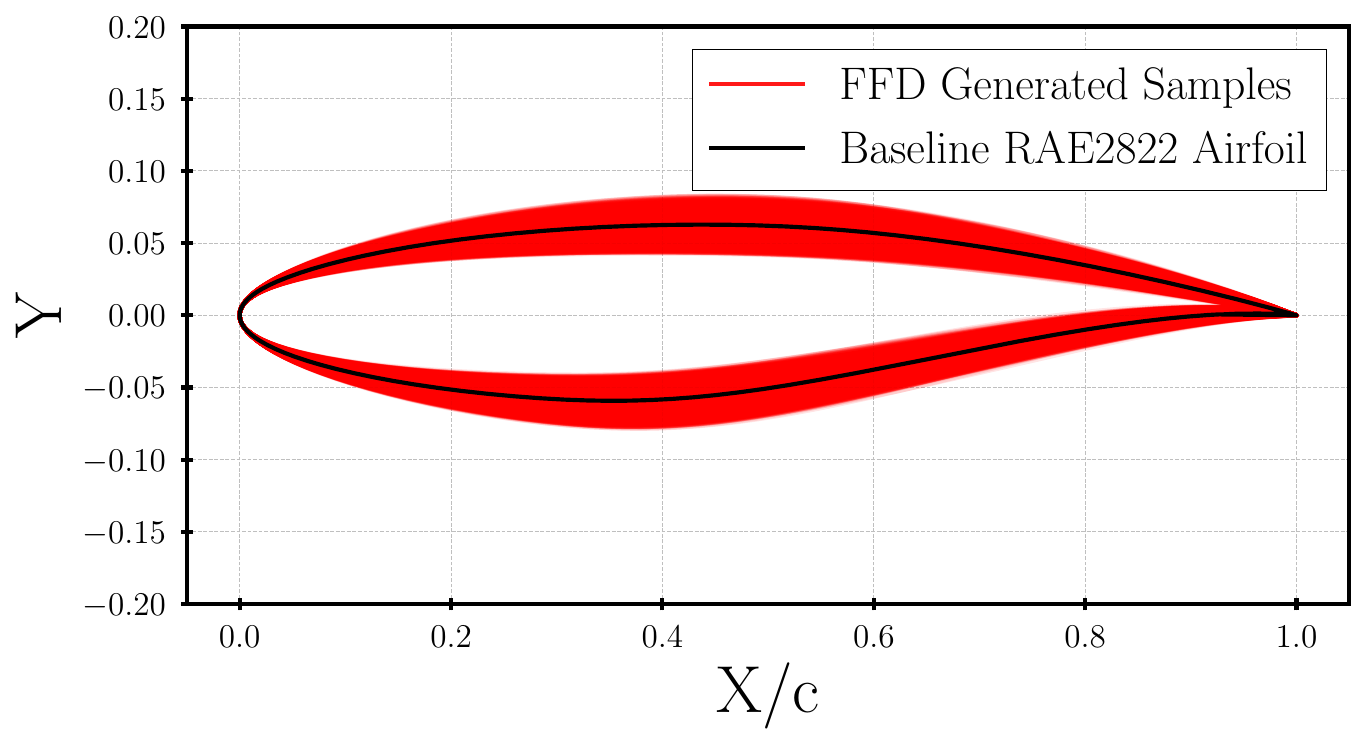}
    \caption{Airfoil shapes generated using the FFD geometric parameterization applied to the RAE2822 airfoil. The red-shaded area represents the variation in airfoil shapes achieved by displacing the control points.}
    \label{fig:generated_samples}
\end{figure}

For this study, the baseline RAE2822 airfoil geometry is enclosed in a $4 \times 2$ FFD box as shown in Fig.~\ref{fig:FFD}. The four corner nodes are kept fixed, while the remaining four nodes are constrained to move only in the vertical direction with a maximum vertical displacement limit of \(\pm 0.03\) times the chord length. The various airfoil shapes generated using the FFD parameterization are shown in Fig.~\ref{fig:generated_samples}.

\begin{figure*}[]
    \centering
    \includegraphics[width=0.85\textwidth]{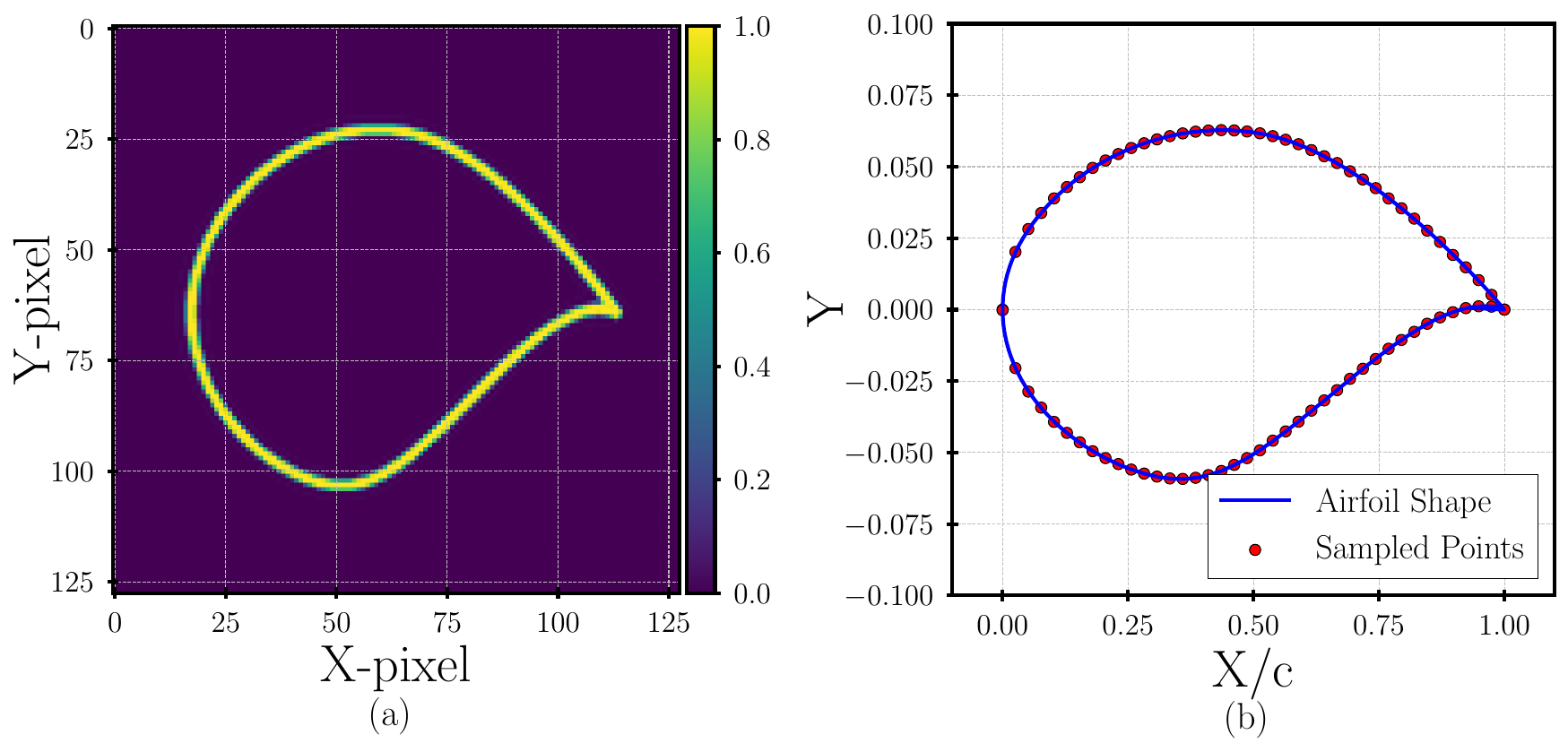}
    \caption{Representation of input and output used for training of CNN-based parameterization network: (a) inverted and normalized greyscale image used as an input, (b) sampled y-coordinates for output.}
    \label{fig:input_output}
\end{figure*}

The input to the CNN-parameterization network is an inverted greyscale image of size \(128 \times 128\), as shown in Fig.~\ref{fig:input_output}(a). The greyscale image is then normalized to obtain pixel values between 0 and 1. Normalization is crucial as it standardizes the input data, improving the training process and convergence of the CNN. A pixel value of 0 represents an area where the airfoil shape does not pass, a value of 1 represents a pixel fully covered by the airfoil, and any value between 0 and 1 indicates partial coverage~\cite{Guo2016ConvolutionalApproximation}. The output values of the parameterization network are the y-coordinates of each airfoil shape sampled at 80 fixed x-coordinate locations as shown in Fig.~\ref{fig:input_output}(b), selected to balance resolution with computational efficiency. These y-coordinates are then centered and normalized to a range of \((-1, 1)\) to ensure compatibility with the activation functions used in the fully connected layers of the network.

It is worth noting that the sampling of y-coordinates along the chord is equidistant, resulting in relatively coarse sampling near the leading edge (LE). The LE is a critical region for aerodynamic performance, and higher-resolution sampling is typically desirable in this area. However, the coarse sampling near the LE does not significantly affect the performance of the CNN-based parameterization network in this study. This is because the movable FFD nodes are positioned away from the LE and closer to the center of the airfoil. Consequently, the generated airfoil designs exhibit minimal shape variation around the LE, as evident in Fig.\ref{fig:FFD} and \ref{fig:generated_samples}. Moreover, the primary objective of the parameterization network is to extract nonlinear shape modes that represent geometric variations. This objective is less sensitive to local discrepancies near the LE compared to optimization problems where precise control over geometry is required. In scenarios where significant shape variations occur near the LE, such as with movable FFD nodes positioned in the LE region, a finer sampling resolution would be necessary to capture these variations accurately.

\subsection{Flow Field Data Set}\label{subsct:field_data_generation}

\begin{figure*}[]
    \centering
    \includegraphics[width=0.85\textwidth]{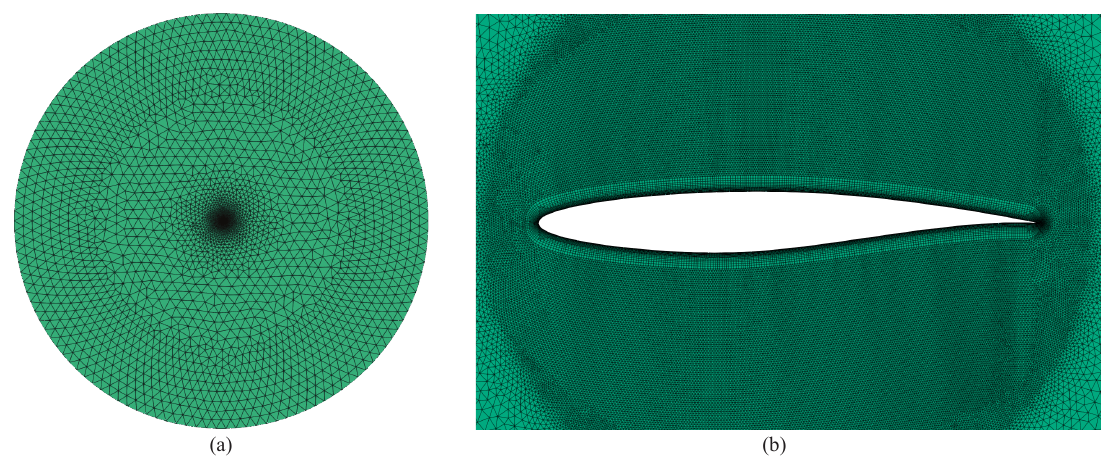}
    \caption{Unstructured grid used for analysis: (a) full domain showing the extent of the flow field, (b) close-up view of the grid around the airfoil.}
    \label{fig:grid_domain}
\end{figure*}

Flow field data around the generated airfoil shapes were obtained using an unstructured grid, demonstrating the capability of our DM-ROM framework to handle such grids directly. For aerodynamic analysis, unstructured grids offer enhanced flexibility for arbitrary shapes, local grid refinement capabilities, and efficiency in grid generation time and effort. Prism layers are generated near the airfoil surface, ensuring that \(y^+ < 1\) is maintained. The final grid selection after the grid independence study contains 954 nodes on the airfoil surface, with a total grid size of 96,913 nodes, as shown in Fig.~\ref{fig:grid_domain}. Details of the grid independence study can be found in the work by Mufti \textit{et al.}~\cite{Mufti2022AInputs,Mufti2023DesignSubspaces}.

RANS-based CFD simulations for the generated airfoil samples were performed under transonic flow conditions using the open-source SU2 code~\cite{economon2016su2}. The free-stream conditions include a Mach number of \(M_\infty = 0.729\), a Reynolds number of \(Re_\infty = 6.5 \times 10^6\), and an angle of attack (\(\alpha\)) sweep ranging from \(0^\circ\) to \(4^\circ\). The Spalart-Allmaras model is chosen for turbulence modeling, while a backward Euler scheme is used to ensure steady flow conditions. A total of 2500 samples are generated from a design space containing 5 design variables (4 FFD nodes and \(\alpha\)). Each sample in our dataset represents a unique airfoil shape and flow conditions, allowing us to generate flow field cases containing shock waves of varying strengths and locations.

\subsection{Performance Metrics}\label{subsct:metrics}

In this study, the pressure coefficient (\(C_p\)) field is chosen as the primary output metric because of its critical role in aircraft design. The \(C_p\) field significantly influences aerodynamic coefficients, making it essential for aero-structural analysis and optimization. While this work focuses on the \(C_p\) field, the framework we propose is versatile enough to be adapted for other aerodynamic fields, such as temperature and velocity distributions. The CNN-parameterization network within our DM-ROM framework is trained solely on airfoil shape data, making it independent of the flow field. This independence significantly reduces the computational cost when adapting the model to predict different output fields, as only the ISOMAP and MLP regression networks would need to be retrained for new flow conditions.

\subsubsection{Field prediction error}\label{subsubsct:rmse}

To evaluate the global accuracy of our model within the flow domain, we utilize the root-mean-squared error (RMSE) as the primary performance metric. Given a testing set consisting of \( n_t \) designs, which were excluded during training and validation, the RMSE is calculated using the following equation:

\begin{equation}\label{eq:field_rms}
E(\mathbf{y}) = \sqrt{\frac{1}{n_t} \sum_{i = 1}^{n_t} \lVert \mathbf{y}_i - \mathbf{\tilde{y}}_i \rVert^2_2}
\end{equation}

In this equation, \( \mathbf{y}_i \) represents the true field solution for the \(i\)-th design in the testing set, and \( \mathbf{\tilde{y}}_i \) denotes the corresponding prediction from our DM-ROM. To facilitate comparison across different models or methodologies, the prediction error is normalized by the standard deviation of the testing dataset, as shown below:

\begin{equation}\label{eq:norm-pred-error}
\widehat{E}(\mathbf{y}) = \frac{E(\mathbf{y})}{\sqrt{\frac{1}{n_t} \sum_{i = 1}^{n_t} \lVert \mathbf{y}_i - \overline{\mathbf{y}}\rVert^2_2}}
\end{equation}

We specifically avoid using relative error metrics for assessing field accuracy, primarily because \( C_p \) values are generally close to the free-stream \( C_p \) value, which is zero. Using relative error in such cases could lead to disproportionately large error values when the denominator is near zero, even if the absolute deviation is minimal. Although introducing a bias term could address this issue, it would introduce scale dependency, which is not desirable~\cite{Duru2022AAirfoils}.

\subsubsection{Shock wave location and strength errors}\label{subsubsct:error_shock}

Prediction errors in the vicinity of shock waves have traditionally been assessed qualitatively. However, some recent studies have sought to quantify these errors using heuristic methods, identifying regions near the shock wave where prediction deviations exceed a defined threshold~\cite{Decker2023ManifoldModeling,Iyengar2022NonlinearDecomposition}. However, these methods implicitly assume that models generally perform very poorly in shock regions, which may not hold for more advanced models. To provide a more rigorous evaluation, we introduce metrics that quantify the accuracy of shock wave predictions in terms of both location and strength.

\begin{figure*}[]
    \centering
    \includegraphics[width=1\textwidth]{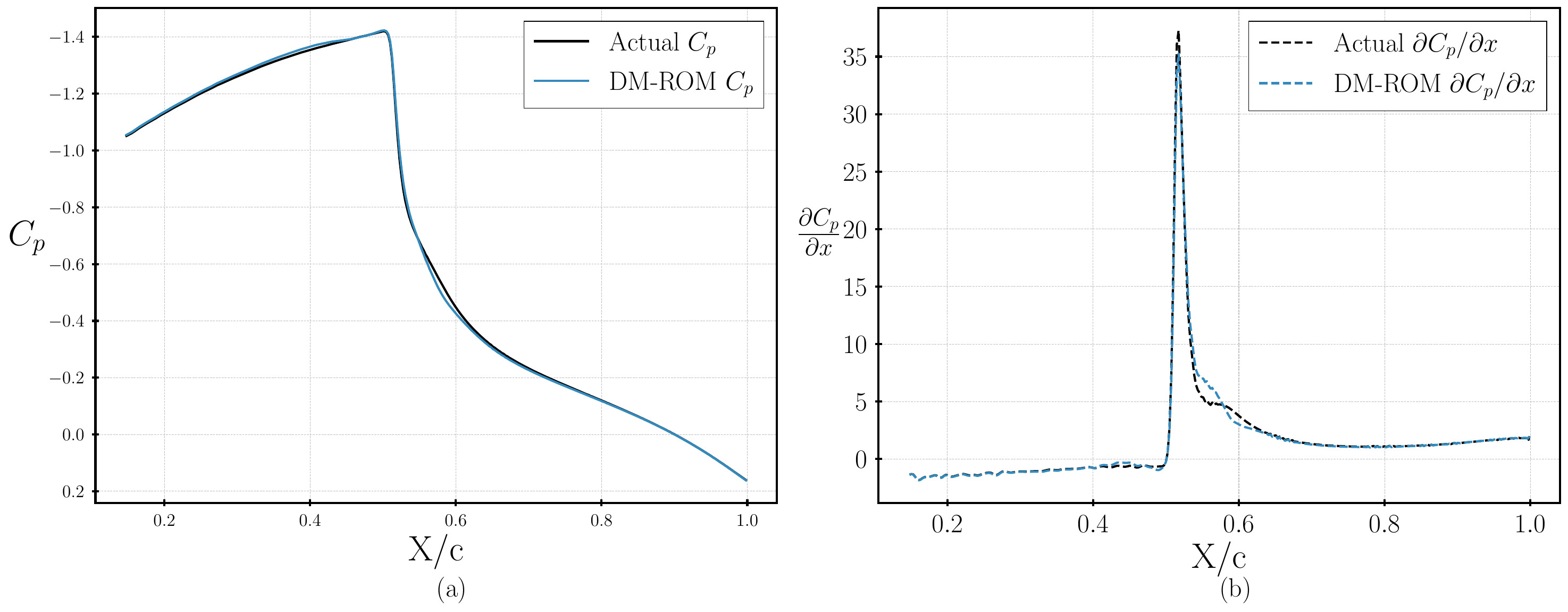}
    \caption{Representation of shock wave on the airfoil top surface (a) actual vs predicted $C_p$, (b) actual vs predicted $C_p$ gradients.}
    \label{fig:shock_representation}
\end{figure*}

We examine the \(C_p\) distribution on the top surface of the airfoil surface where shock waves are likely to form, as illustrated in Fig.~\ref{fig:shock_representation}. A sudden spike in the \(C_p\) gradient along this surface identifies a shock wave. The shock wave location, \(x^{s}\), is determined by identifying the index \(i\) where:

\begin{equation}
\frac{\left( \frac{\partial C_p}{\partial x} \right)_i}{ \left( \frac{\partial C_p}{\partial x} \right)_{max}} > \gamma
\end{equation}

Here, \(\gamma\) is a threshold parameter, set to \(\gamma = 0.1\) in this study based on engineering judgment and analysis of the actual flow field data. The error in shock wave location prediction is then computed by comparing the actual \(x^{s}\) with the predicted \(\tilde{x}^{s}\). The shock location error for the test set is then calculated as:

\begin{equation}\label{eq:mape_location}
    \widehat{E}_{sl}(x^{s}(i)) = \frac{1}{n_{t}} \sum_{j=1}^{n_t}  \Bigg{[} \frac{|x^{s}(i) - \Tilde{x}^{s}(i)|}{|x^{s}(i)|} \Bigg{]}_{j} 
\end{equation}

Additionally, the shock wave strength, \(\delta C_p\), can be determined by identifying the end location of the shock wave, \(x^{e}\), where the gradient falls below the threshold after the shock start index:

\begin{equation}
\frac{\left( \frac{\partial C_p}{\partial x} \right)_{i^*}}{ \left( \frac{\partial C_p}{\partial x} \right)_{max}} < \gamma
\end{equation}

The shock strength is calculated as the difference in \(C_p\) values between \(x^s\) and \(x^e\). The error in predicting the shock strength is given by:

\begin{equation}\label{eq:mape_strength}
    \widehat{E}_{ss}(\delta C_{p}) = \frac{1}{n_{t} }\sum_{j=1}^{n_t}  \Bigg{[}\frac{|\delta {C_p} - \Tilde{\delta C_p}|}{|\delta {C_p}|}  \Bigg{]}_{j} 
\end{equation}
where \(\delta C_p = | C_p(x^s) - C_p(x^e)|\). In calculating the shock wave strength and location error across the entire testing dataset, we exclude cases devoid of a shock wave to ensure the error metric is not unduly influenced. The error metrics introduced in this section will be further leveraged in section~\ref{sct: results} to assess the precision of our proposed methodology in forecasting shock waves with diverse strengths and positions.

\section{Results and Discussion}\label{sct: results}
This section presents the outcomes of applying our DM-ROM methodology to the RAE2822 airfoil transonic flow field test case. The generated datasets are divided into training, validation, and testing sets. We randomly select 500 samples as the testing set from the total dataset. Out of the remaining 2000 samples, 20\% are used as the validation set, with the rest reserved for training. Later in this section, we explore the effect of varying the number of training samples on the performance of the DM-ROM methodology and compare it with reference ROM methodologies. In these variations, the testing set remains consistent.

\subsection{Training of CNN-based Parameterization Network}\label{subsct:cnn_training}

The CNN-based parameterization networks were trained using the \textit{TensorFlow} framework. A learning rate scheduler was employed to reduce the learning rate when a plateau in learning was detected. Various hyperparameter values were tested, including training batch size, initial learning rate, learning rate scheduler reduction factor, patience, and minimum learning rate. While the hyperparameter selection study is not shown here for brevity, the final hyperparameters selected for training the CNN networks are summarized in Table~\ref{tab:cnn_hyperparameters}.

\begin{table}[]
\caption{\label{tab:cnn_hyperparameters} Hyperparameters selected for training CNN-based parameterization networks.}
\begin{ruledtabular}
\begin{tabular}{@{}lc@{}}
Hyperparameters & Values \\
\hline
Training batch size & 16 \\
Initial learning rate & 1 $\times$ $10^{-4}$ \\
\begin{tabular}[c]{@{}l@{}}Learning rate scheduler\end{tabular} &  \\
\multicolumn{1}{r}{Factor} & 0.7 \\
\multicolumn{1}{r}{Patience} & 20 \\
\multicolumn{1}{r}{Minimum learning rate} & 1 $\times$ $10^{-7}$
\end{tabular}
\end{ruledtabular}
\end{table}

\begin{figure}[]
    \centering
    \includegraphics[width=0.65\textwidth]{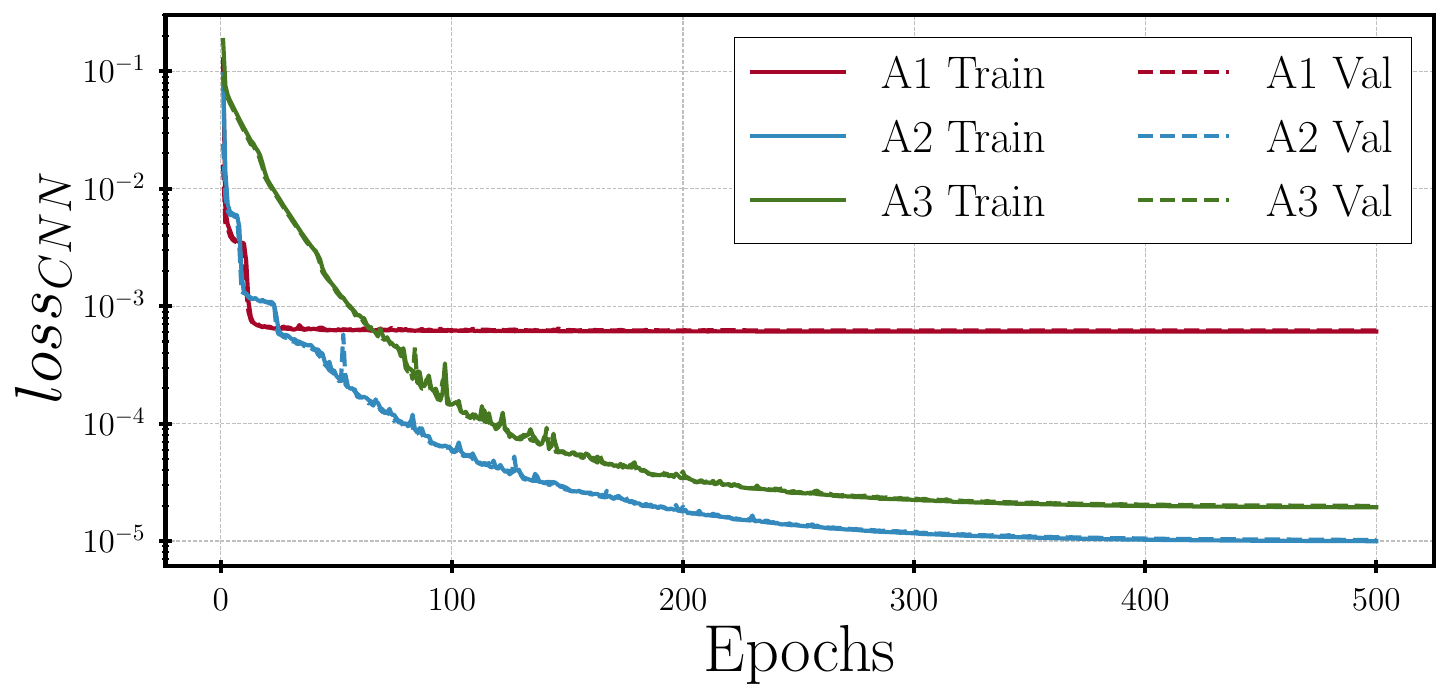}
    \caption{Training and validation loss curves for CNN-based parameterization networks.}
    \label{fig:loss_curve}
\end{figure}

Using the selected hyperparameters, the three CNN architectures were trained on an NVIDIA Tesla V100 node. The number of nonlinear shape parameters \(p\) to be extracted was set to four for all networks to match the number of FFD nodes used. However, the CNN-parameterization technique offers flexibility to vary the number of shape parameters by adjusting the number of neurons in the parameter layer. The loss function curves for the training and validation sets for the three CNN architectures are shown in Fig.~\ref{fig:loss_curve}. The training was conducted for 500 epochs in all cases. Initially, both training and validation losses dropped, but the curves for the A1 architecture settled at a higher \(loss_{CNN}\) value. In contrast, the A2 and A3 architectures continued to show a gradual reduction in loss until convergence. Table~\ref{tab:cnn_selection_results} presents the final loss function values for the training and validation sets, along with the time required to complete 500 epochs. The table reveals that while both the A2 and A3 networks achieved similar training and validation loss values, the computational time required for the A3 network was more than twice that of the A2 network.

\begin{table}[h]
\caption{\label{tab:cnn_selection_results} Final loss values and training time for different CNN-based parameterization networks.}
\begin{ruledtabular}
\begin{tabular}{@{}lccc@{}}
Architecture & \(loss_{CNN}\) Train & \(loss_{CNN}\) Val & Training time (min)$^{\footnotemark[1]}$ \\ 
\hline
A1 & $6.08 \times 10^{-4}$ & $6.21 \times 10^{-4}$ & 13.27 \\
A2 & $1.46 \times 10^{-5}$ & $1.52 \times 10^{-5}$ & 32.81 \\
A3 & $1.93 \times 10^{-5}$ & $1.95 \times 10^{-5}$ & 84.55 \\ 
\end{tabular}
\end{ruledtabular}
\footnotetext[1]{Model training was performed using an NVIDIA Tesla V100 GPU node with 2 GPUs and 24 Intel Xeon Gold 6226 CPUs. Training time for 500 epochs.}
\end{table}

\begin{figure*}[]
    \centering
    \includegraphics[width=0.85\textwidth]{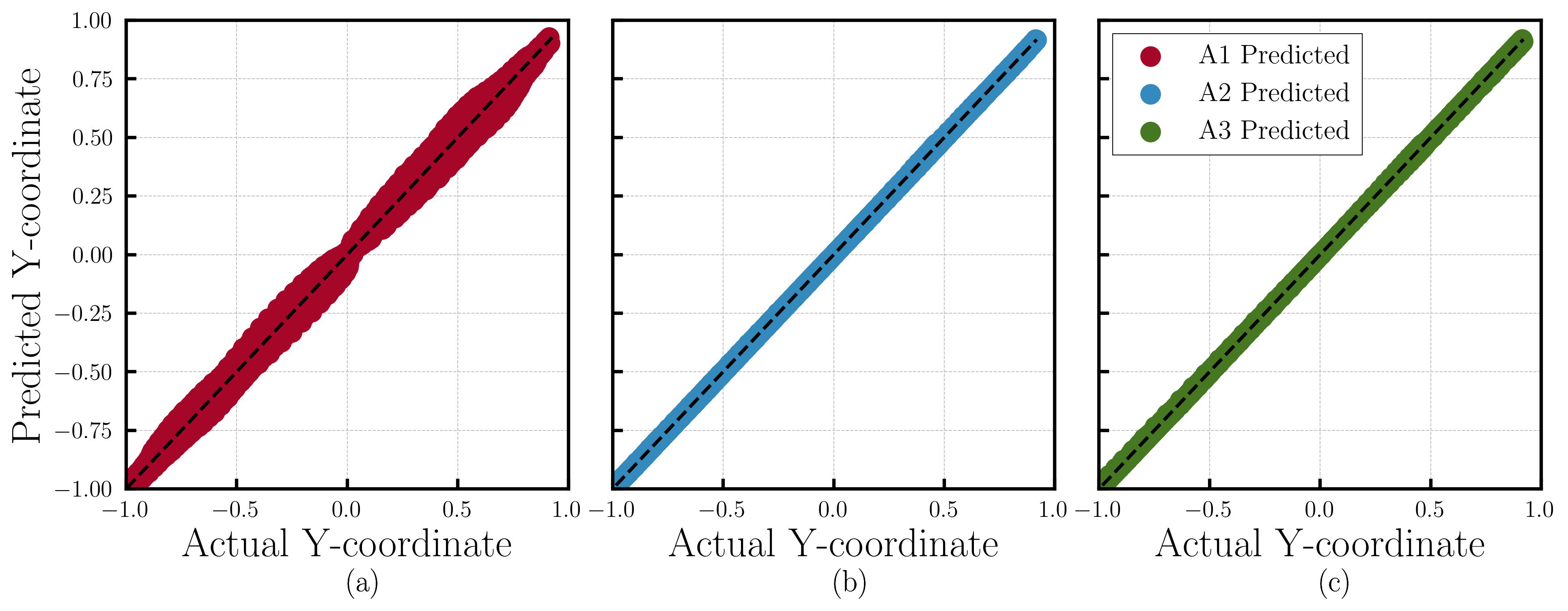}
    \caption{Actual vs predicted y-coordinates for airfoil shapes in the testing set: (a) A1 network, (b) A2 network, and (c) A3 network.}
    \label{fig:actual_predicted}
\end{figure*}
\begin{figure*}[]
    \centering
    \includegraphics[width=1\textwidth]{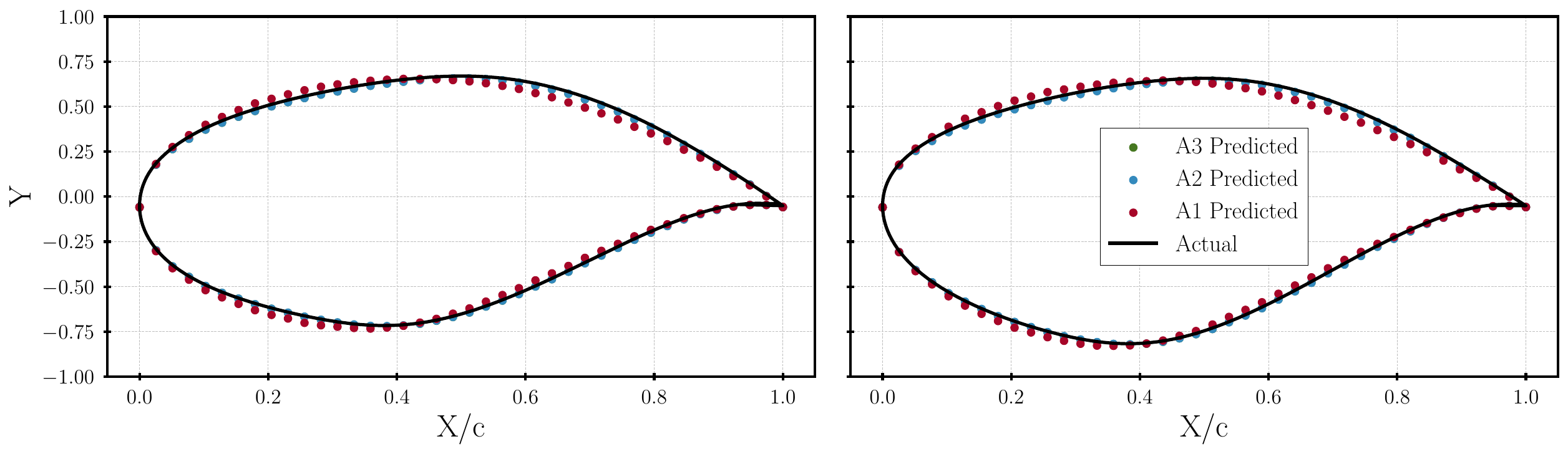}
    \caption{Comparison of actual airfoil shapes with those reconstructed from the predicted nonlinear shape parameters for two randomly selected samples from the testing set.}
    \label{fig:airfoil_shapes_closed_loop}
\end{figure*}

The actual versus predicted plots for the y-coordinates of all the airfoil shapes in the testing set are shown in Fig.~\ref{fig:actual_predicted}. The A1 network exhibits slight deviations from the mean fit line, with some under- and over-predictions. In contrast, both the A2 and A3 networks demonstrate a good fit for all predicted coordinates. The nonlinear shape modes extracted by the network are nonintuitive and do not have a direct physical interpretation. Instead of plotting these abstract modes, we reconstruct the airfoil shapes using the nonlinear parameters and the fully connected part of the network, which decodes them back to physical coordinates. Figure~\ref{fig:airfoil_shapes_closed_loop} compares the shapes of two randomly selected airfoils from the testing set as reconstructed using the four nonlinear shape parameters from the three architectures. The airfoil shapes recovered using the predictions from the A2 and A3 networks closely match the actual airfoil shapes. This comparison indicates that while both the A2 and A3 networks offer strong parameterization capabilities, the A2 network is significantly more computationally efficient and is therefore selected for further analysis in this study.

It is important to highlight that the CNN-based parameterization network only requires geometric shape information to train, without needing flow field data. This makes the dataset much cheaper and faster to generate, and a large number of shape samples can be created for complex aerodynamic geometries, ensuring robust training even for intricate designs.

\subsection{Selection of ISOMAP manifold dimension and training of MLP regression network}\label{subsct: mlp_training}

\begin{figure}[]
    \centering
    \includegraphics[width=0.65\textwidth]{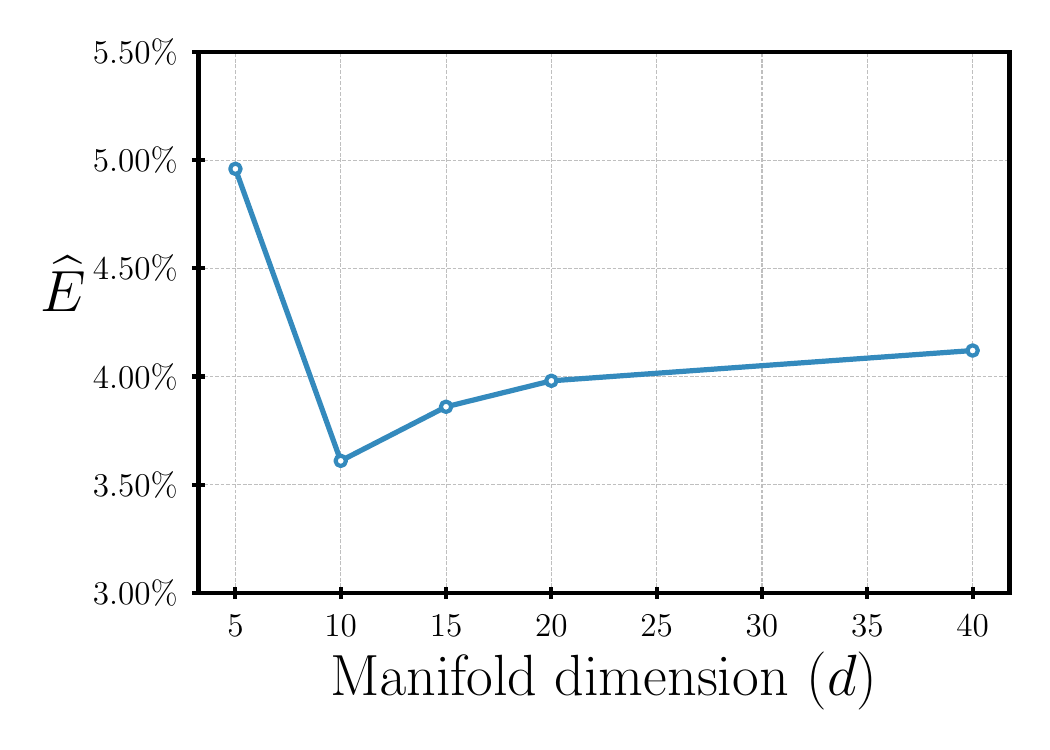}
    \caption{Effect of manifold dimension (\(d\)) on the prediction accuracy of DM-ROM.}
    \label{fig:isomap_effect}
\end{figure}

The ISOMAP manifold learning problem is not well-posed because the intrinsic dimensionality \(d\) of the dataset is not known \textit{a priori}. One approach is to assume that the manifold dimension is approximately equal to the number of input parameters varied to obtain the output data~\cite{franz2016reduced}. However, in this study, we vary the number of manifold dimensions to observe their effect on the field prediction accuracy of the DM-ROM and determine the manifold dimension that yields the best results. Figure~\ref{fig:isomap_effect} shows the variation in field prediction error \(\widehat{E}\) with the manifold dimension \(d\). For very low manifold dimension, the field prediction error is around 5\%. As \(d\) increases to 10, a noticeable drop in prediction error occurs. Beyond a manifold dimension of 10, a slight increase in \(\widehat{E}\) is observed. These results suggest that the intrinsic dimensionality of the dataset is around 10, and increasing the manifold dimension further may lead to overfitting, resulting in a slight increase in the field prediction error. Based on these results, a manifold dimension of \(d = 10\) is selected for subsequent analyses.

The impact of the depth and width of different MLP architectures on the predictive accuracy of the DM-ROM is evaluated by comparing the field prediction error and training time for different MLP regression networks listed in Table~\ref{tab:mlp_selection_results}. As the complexity of the MLP networks increases, either by adding layers or increasing the number of nodes per layer, a reduction in field prediction error is observed, alongside an increase in computational time. The M6 network, consisting of 8 fully connected layers with 800 neurons per layer, achieves the best predictive accuracy of 3.61\%. However, further increasing the complexity to the M7 network (10 layers with 1000 neurons each) leads to a slight increase in prediction error due to increase in number of model parameters, accompanied by a significant increase in training time. Given the balance between cost and prediction accuracy, the M6 network is selected for further analysis in this study. 

\begin{table}[]
\caption{\label{tab:mlp_selection_results} Summary of field prediction error and training time for different MLP regression networks.}
\begin{ruledtabular}
\begin{tabular}{@{}lcc@{}}
Architecture &  \( \widehat{E}\) &  Training time (min)$^{\footnotemark[1]}$ \\ 
\hline
M1 (10 $\times$ 100) & 4.11\% &  6.27 \\
M2 (10 $\times$ 200)& 3.90\% & 6.45 \\
M3 (5 $\times$ 500) & 3.76\% & 6.76 \\
M4 (8 $\times$ 400) & 3.95\% & 6.53 \\
M5 (6 $\times$ 600) & 3.63\% & 7.62 \\
M6 (8 $\times$ 800) & 3.61\% & 8.64 \\
M7 (10 $\times$ 1000) & 3.79\% & 10.3 \\
\end{tabular}
\end{ruledtabular}
\footnotetext[1]{Model training was conducted using an NVIDIA Tesla V100 GPU node containing 2 GPUs and 24 Intel Xeon Gold 6226 CPUs.}
\end{table}

\subsection{Prediction of Flow Field}\label{subsct:flowfield_prediction}

\begin{figure*}[]
    \centering
    \includegraphics[width=1\textwidth]{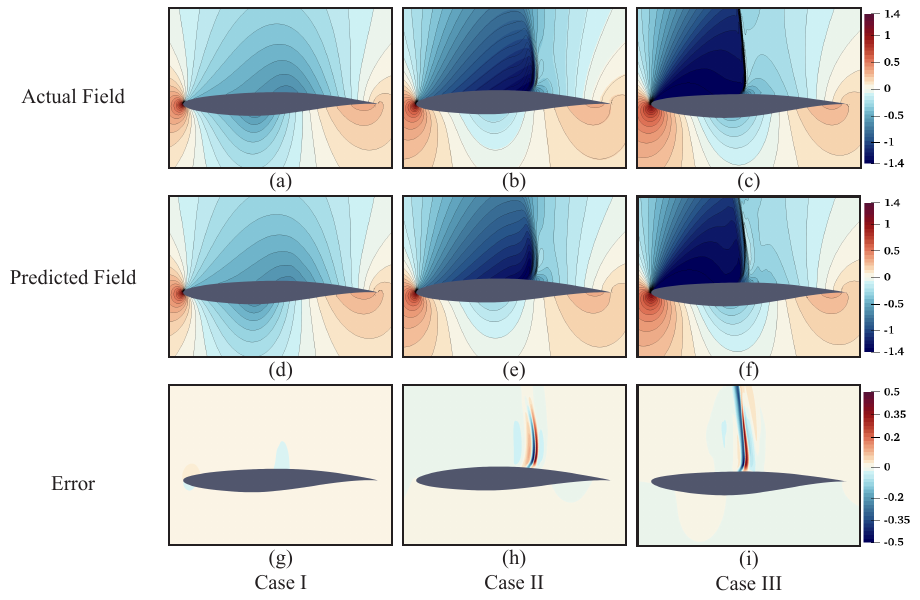}
    \caption{Contour plots for three different cases: (a)-(c) actual $C_p$ field from CFD, (d)-(f) predicted $\tilde{C}_p$ field from DM-ROM, and (g)-(i) error $C_p - \tilde{C}_p$ field.}
    \label{fig:flowfield_model}
\end{figure*}

We evaluate the predictive performance of the DM-ROM by analyzing the flow field predictions for various airfoil designs and angles of attack. Figure~\ref{fig:flowfield_model} presents the actual CFD flow field, the predicted flow field, and the prediction error fields, computed as \(C_p - \tilde{C}_p\), for three different cases selected from the testing data set. These cases are representative of flow fields containing shock waves of varying strengths and locations.

In case I, where no shock wave is observed, the DM-ROM achieves high accuracy in predicting the flow field. Similarly, in case II, which contains a normal shock wave, the DM-ROM continues to perform well, with only minimal prediction errors observed. The errors are localized in narrow bands immediately downstream of the shock wave. For case III, which features a stronger shock wave and significant shock-boundary layer interaction, the error bands become more pronounced near the shock wave region. However, even in this challenging case, the magnitude of the prediction errors remains relatively low, demonstrating that the DM-ROM can handle complex shock-induced flow features. Overall, the DM-ROM exhibits excellent performance across all cases, accurately capturing both weak and strong shock waves.

\begin{figure*}[]
    \centering
    \includegraphics[width=1\textwidth]{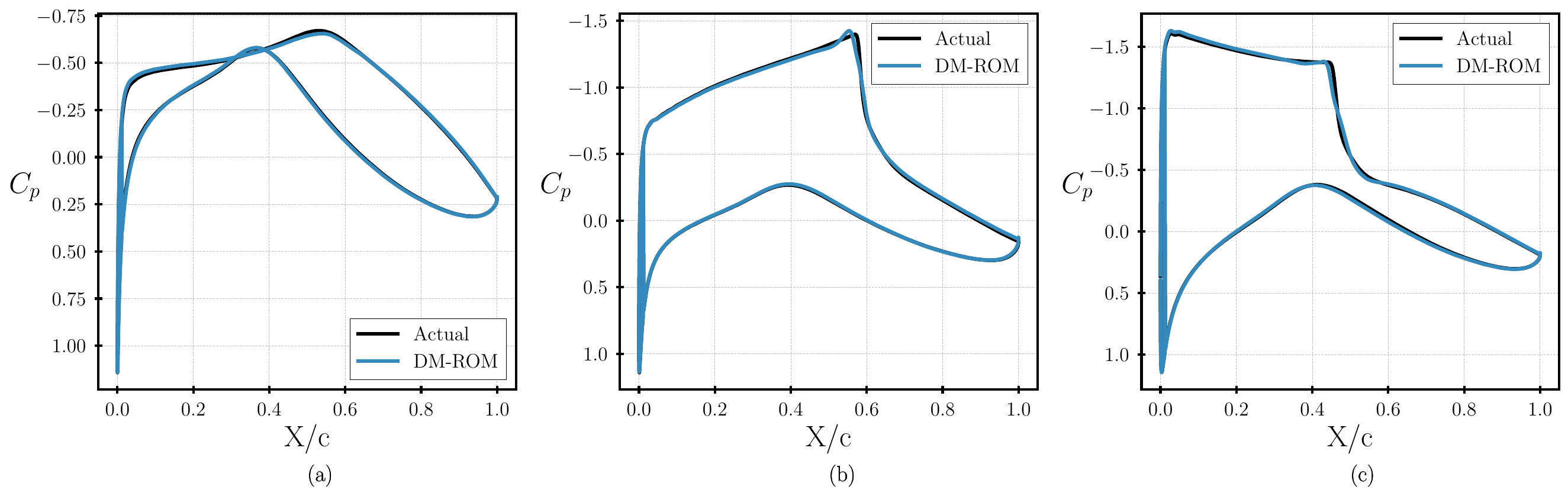}
    \caption{Airfoil surface pressure distribution $C_p$ for the three cases shown in Fig.~\ref{fig:flowfield_model}: (a) case I, (b) case II, and (c) case III. The shock wave locations are indicated by a sharp rise in $C_p$ on the airfoil surface.}
    \label{fig:Cp_surface}
\end{figure*}

To further assess the DM-ROM's capability to accurately predict surface pressure distributions and capture shock wave positions, we plot the distribution of \(C_p\) along the airfoil surface for the three cases presented in Fig.~\ref{fig:flowfield_model}. Figure~\ref{fig:Cp_surface} shows the variation of \(C_p\) across the airfoil surface for cases I, II, and III. In cases II and III, the shock waves manifest as sudden jumps in the pressure distribution. The DM-ROM successfully predicts these shock waves, as evidenced by the close agreement between the actual and predicted pressure distributions. This suggests that the DM-ROM accurately captures the complex aerodynamic behavior, including shock phenomena, across the testing dataset.

\subsection{Comparison with Other ROM Techniques}\label{subsct:rom_comparison}

We compare the prediction accuracy of the proposed DM-ROM methodology with two widely used ROM techniques from the literature: a linear POD-ROM and a nonlinear ISOMAP-ROM. Both reference ROM techniques directly use FFD variables as input and employ Kriging for regression. The output dimension reduction is performed using POD and ISOMAP for the respective techniques. To ensure a fair comparison, the number of nonlinear shape parameters extracted by the CNN-parameterization network in DM-ROM was set equal to the number of FFD variables used by the reference techniques. Moreover, the number of low-dimensional output latent variables \(d\) was kept constant for all three ROM techniques. For POD-ROM, the number of modes (\(d\)) was selected based on the relative information criterion (RIC)~\cite{pinnau2008model}, targeting 99\% of variability in the field data. The same number of modes was then used for ISOMAP-ROM and DM-ROM to maintain consistency in the comparison.

\subsubsection{Effect of number of training samples}\label{subsct:training_samples_effect}

\begin{figure}[]
    \centering
    \includegraphics[width=0.65\textwidth]{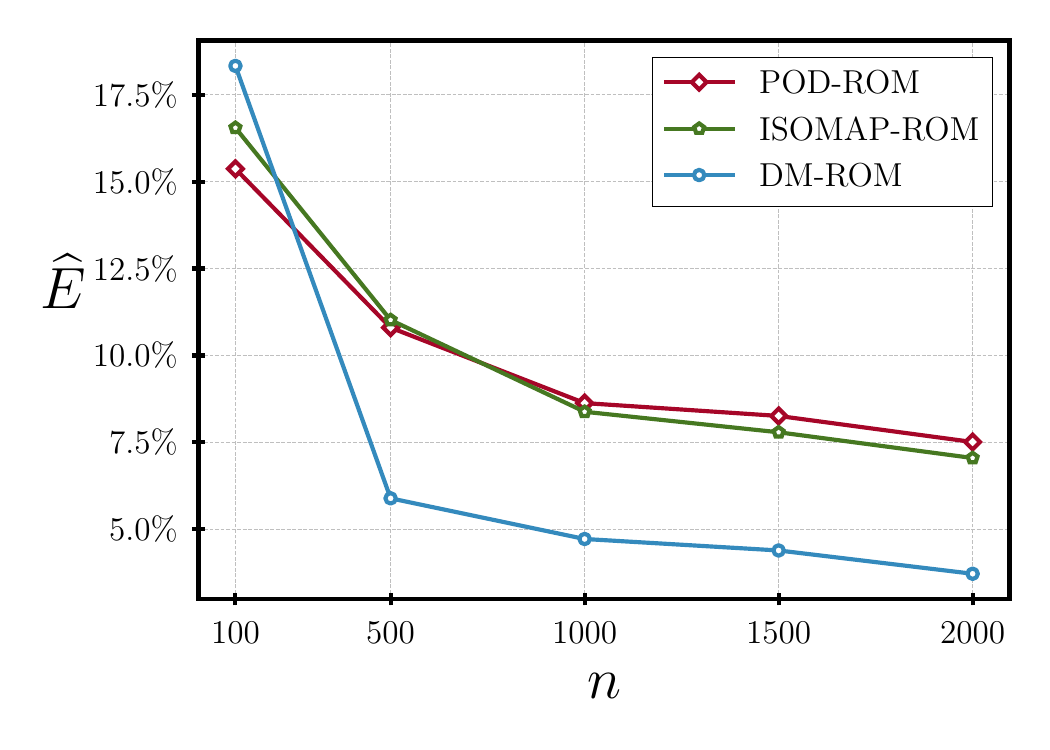}
    \caption{Comparison of field prediction error as a function of training samples for POD-ROM, ISOMAP-ROM, and DM-ROM.}
    \label{fig:training_samples_effect}
\end{figure}

In real-world design applications, the number of training samples available to develop a ROM is often constrained by computational budget limitations. Therefore, it is essential that a robust ROM methodology performs well across different sample sizes. We evaluated the predictive accuracy of POD-ROM, ISOMAP-ROM, and DM-ROM for varying training sample sizes, as shown in Fig.~\ref{fig:training_samples_effect}.

At lower sample sizes, DM-ROM exhibits higher prediction errors compared to both POD-ROM and ISOMAP-ROM. This is because the CNN-parameterization network in DM-ROM requires a sufficient amount of training data to effectively learn nonlinear shape parameters. Similarly, the MLP regression network, being fully connected, needs a larger dataset to accurately predict the low-dimensional output variables. On the other hand, POD-ROM and ISOMAP-ROM, which directly use FFD variables and Kriging for regression, show better resilience to sparse data conditions.

As the number of training samples increases from 100 to 500, we observe a significant reduction in prediction errors across all ROM techniques, with DM-ROM showing the most pronounced improvement. This indicates that with sufficient training data, the CNN-parameterization and MLP regression networks in DM-ROM can effectively capture complex relationships between the input variables and the flow field outputs. As the number of training samples is further increased, the prediction errors continue to decrease for all methods. When using the full training set of 2000 samples, DM-ROM achieves approximately half the prediction error of POD-ROM and ISOMAP-ROM, as shown in Table~\ref{tab:training_sample_effect}. These results highlight the superior performance of DM-ROM in predicting the complete flow field with greater accuracy, especially when sufficient training data is available.

\begin{table}
\caption{\label{tab:training_sample_effect} Comparison of field prediction errors between POD-ROM, ISOMAP-ROM, and DM-ROM for different training sample sizes.}
\begin{ruledtabular}
\begin{tabular}{cccccc}
{$n$} & Modes (\(d\)) & \multicolumn{3}{c}{\(\widehat{E}\)}  \\
& & POD-ROM & ISOMAP-ROM & DM-ROM & \\
\hline
100  & 18 & 15.37\%  & 16.55\%  & 18.33\%  \\
500  & 21 & 10.80\%    & 11.02\%    & 5.89\%   \\
1000 & 22 & 8.63\%     & 8.38\%     & 4.72\%   \\
1500 & 23 & 8.26\%    & 7.79\%     & 4.39\%  \\
2000 & 23 & 7.51\%    & 7.05\%     & 3.72\%   \\
\end{tabular}
\end{ruledtabular}
\end{table}

\subsubsection{Prediction of Shock Waves in the Flow Field}\label{subsct:shock_errors}

In this section, we compare the performance of the DM-ROM technique with reference ROM techniques in predicting shock waves across varying training sample sizes. The ability to predict shock waves accurately is critical, as these are highly nonlinear flow features that have a significant impact on the aerodynamic performance of airfoils, particularly in transonic regimes.

\begin{figure*}[]
    \centering
    \includegraphics[width=1\textwidth]{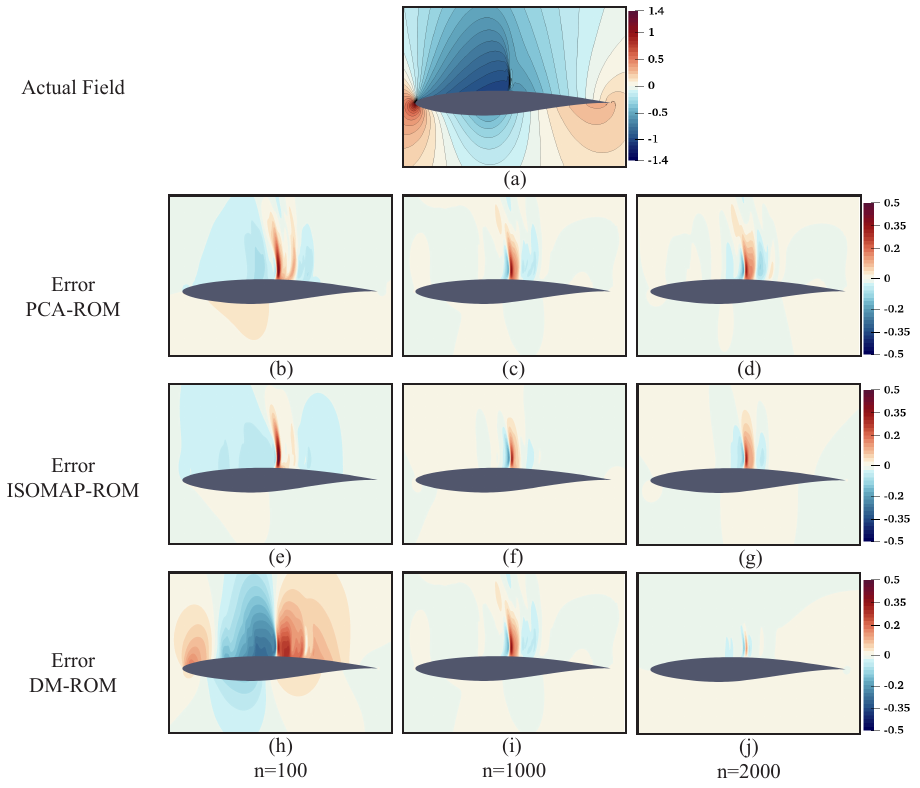}
    \caption{Contour plots for various training sample sizes for weak shock case: (a) CFD $C_p$ field, (b)-(d) POD-ROM error $C_p - \tilde{C_p}$ field, (e)-(g) ISOMAP-ROM error $C_p - \tilde{C_p}$ field, (h)-(j) DM-ROM error $C_p - \tilde{C_p}$ field.}
    \label{fig:flowfield_weak_shock}
\end{figure*}

Figure~\ref{fig:flowfield_weak_shock} represents a scenario where a relatively weak shock wave forms on the upper surface of the airfoil. For small sample sizes (\(n=100\)), we observe that all three ROM techniques—POD-ROM, ISOMAP-ROM, and DM-ROM—struggle to accurately capture the flow field, particularly in regions near the shock wave. DM-ROM exhibits the largest error contours, while POD-ROM and ISOMAP-ROM perform better with smaller errors. This can be attributed to the fact that DM-ROM, with its CNN-parameterization and MLP-regression networks, requires a larger amount of training data to effectively capture the nonlinear relationships between the airfoil shape and the flow field. POD-ROM, which leverages the linear POD decomposition and Kriging regression, is less data-intensive and is better suited for small datasets, as it can generate a reasonable low-dimensional approximation of the field using fewer samples. ISOMAP-ROM, while nonlinear in its dimensionality reduction technique, also benefits from the direct use of linear FFD variables, allowing it to perform reasonably well with sparse data.

As the number of training samples increases to \(n=1000\), we see a significant reduction in the error contours for all ROM techniques. The error zones near the shock wave shrink, particularly for DM-ROM, which starts to leverage its ability to model complex nonlinear relationships more effectively. The CNN-parameterization network is now better trained to capture nonlinear shape variations, and the MLP-regression network is able to accurately map the learned modes to the flow field. In contrast, POD-ROM and ISOMAP-ROM exhibit slower improvements as they lack the flexibility to capture complex, higher-order interactions present in the flow field. 

\begin{figure*}[]
    \centering
    \includegraphics[width=1\textwidth]{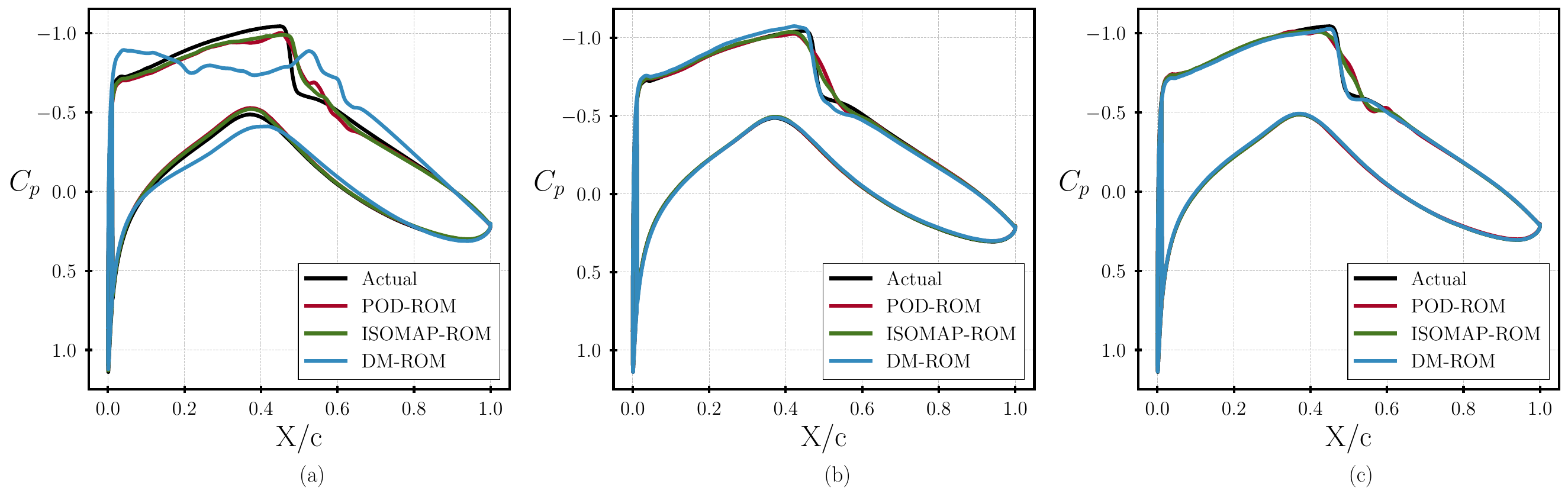}
    \caption{Variation of $C_p$ on the airfoil surface for weak shock case and different training sizes: (a) $n = 100$, (b) $n = 1000$, and, (c) $n = 2000$.}
    \label{fig:Cp_shock_weak}
\end{figure*}

When the full dataset of \(n=2000\) is used, DM-ROM demonstrates superior performance, with error contours in the shock region significantly smaller than those observed for POD-ROM and ISOMAP-ROM. As shown in Figure~\ref{fig:Cp_shock_weak}, DM-ROM captures the \(C_p\) distribution on the airfoil surface with high accuracy, even in the vicinity of the shock wave. In contrast, POD-ROM and ISOMAP-ROM still struggle to predict the shock wave's strength and location accurately, particularly under highly nonlinear flow conditions.

\begin{figure*}[]
    \centering
    \includegraphics[width=1\textwidth]{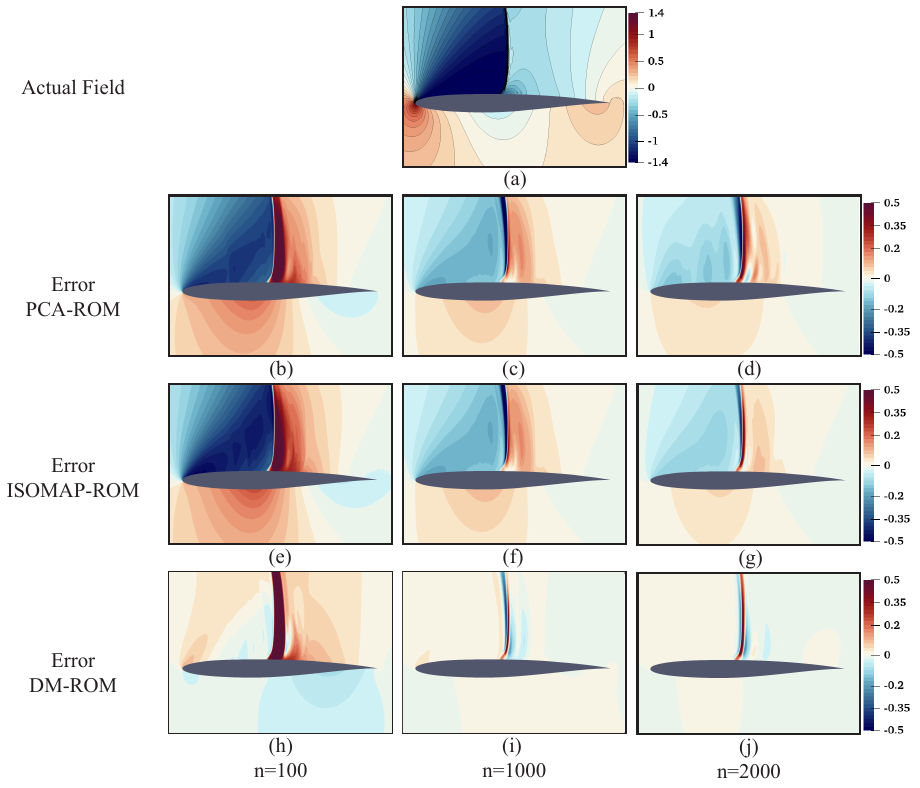}
    \caption{Contour plots for various training sample sizes for strong shock case: (a) CFD $C_p$ field, (b)-(d) POD-ROM error $C_p - \tilde{C_p}$ field, (e)-(g) ISOMAP-ROM error $C_p - \tilde{C_p}$ field, (h)-(j) DM-ROM error $C_p - \tilde{C_p}$ field.}
    \label{fig:flowfield_strong_shock}
\end{figure*}

For a strong shock case, depicted in Figure~\ref{fig:flowfield_strong_shock}, the challenge of accurately predicting the shock wave becomes more apparent. At \(n=100\), all ROM techniques exhibit large error zones, particularly around the shock wave. The error contours are largest for DM-ROM, which reflects its higher dependency on larger datasets to learn the nonlinearities of the shock wave. Despite these initial struggles, as the number of samples increases to \(n=1000\) and \(n=2000\), DM-ROM shows significant improvements in predicting both the location and strength of the shock wave. The error zones shrink substantially, and DM-ROM’s ability to capture the abrupt changes in the flow field is evident. 

\begin{figure*}[]
    \centering
    \includegraphics[width=1\textwidth]{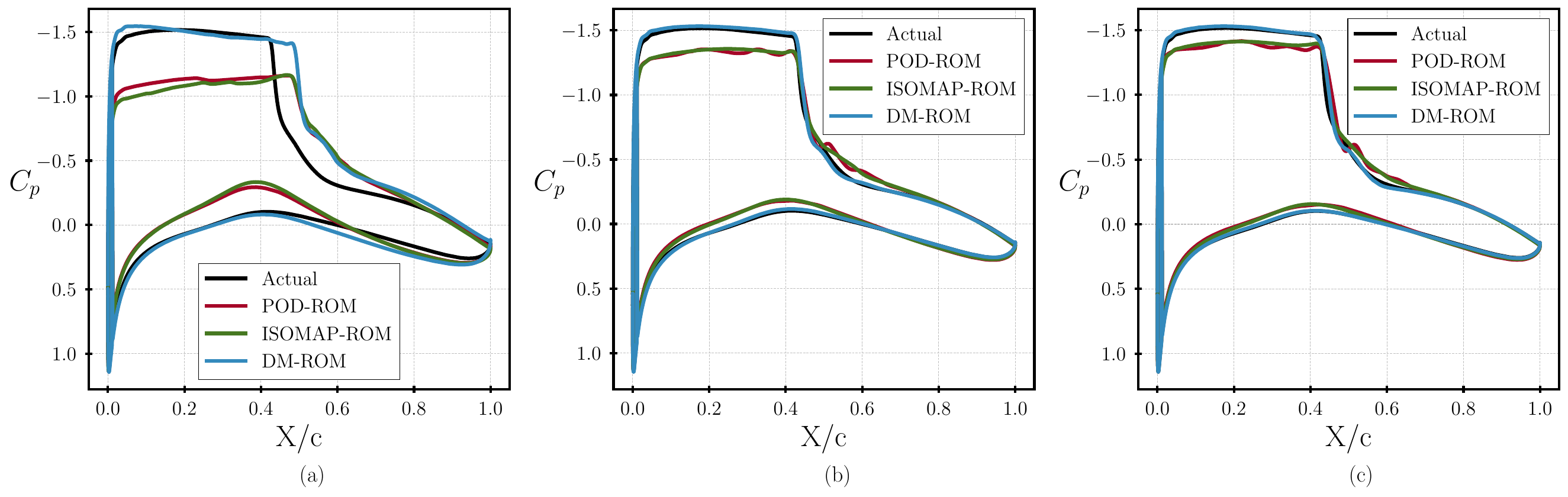}
    \caption{Variation of $C_p$ on the airfoil surface for strong shock case and different training sizes: (a) $n = 100$, (b) $n = 1000$, and, (c) $n = 2000$.}
    \label{fig:Cp_shock_strong}
\end{figure*}

On the other hand, POD-ROM and ISOMAP-ROM continue to display significant error regions even when trained on the full dataset. As shown in Figure~\ref{fig:Cp_shock_strong}, both POD-ROM and ISOMAP-ROM struggle to accurately predict the sudden jump in \(C_p\) associated with the shock wave, particularly under strong shock conditions. In contrast, DM-ROM’s nonlinear framework allows it to match the actual \(C_p\) distribution much more closely.

\begin{figure*}[]
    \centering
    \includegraphics[width=1\textwidth]{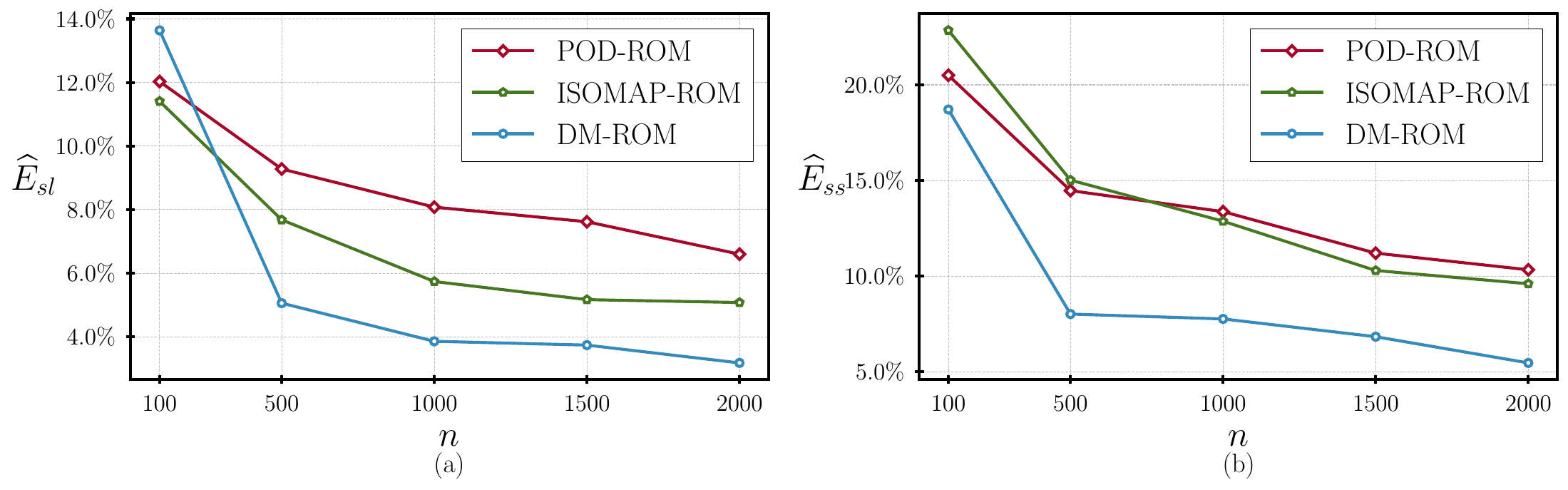}
    \caption{Comparison of shock wave prediction errors with the number of training samples for POD-ROM, ISOMAP-ROM, and DM-ROM: (a) shock wave strength error, and (b) shock wave location error.}
    \label{fig:error_shock_training}
\end{figure*}

Figures~\ref{fig:error_shock_training} show the shock wave location error (\( \widehat{E}_{sl} \)) and shock wave strength error (\( \widehat{E}_{ss} \)) for the three ROM techniques as a function of the number of training samples. At \(n=100\), POD-ROM performs better than both ISOMAP-ROM and DM-ROM in predicting the shock wave location and strength. This is because POD-ROM efficiently captures broad flow features, including approximate shock wave locations, even with limited training data. However, as the number of training samples increases, DM-ROM begins to outperform both reference ROM techniques. By \(n=2000\), DM-ROM achieves roughly half the prediction error of POD-ROM and ISOMAP-ROM in terms of both shock wave location and strength, as shown in Table~\ref{tab:training_sample_effect_shocks}. These results highlight the ability of DM-ROM to model complex nonlinear flow features, such as shock wave interactions, more effectively as sufficient training data becomes available. 

\begin{table*}
\caption{\label{tab:training_sample_effect_shocks} Comparison of shock wave location and strength prediction errors between POD-ROM, ISOMAP-ROM, and DM-ROM for different training sample sizes.}
\begin{ruledtabular}
\begin{tabular}{ccccccc}
 {$n$} &\multicolumn{3}{c}{\(\widehat{E}_{sl}\)} & \multicolumn{3}{c}{\(\widehat{E}_{ss}\)}\\
 & POD-ROM & ISOMAP-ROM & DM-ROM & POD-ROM & ISOMAP-ROM & DM-ROM \\
\hline
100  & 12.03\% & 11.41\% & 13.64\% & 20.50\% & 22.85\% & 18.71\% \\
500  & 9.28\%  & 7.68\%  & 5.06\%  & 14.47\% & 15.02\% & 8.01\%  \\
1000 & 8.08\%  & 5.74\%  & 3.86\%  & 13.37\% & 12.87\% & 7.76\%  \\
1500 & 7.62\%  & 5.17\%  & 3.74\%  & 11.20\% & 10.29\% & 6.83\%  \\
2000 & 6.60\%  & 5.08\%  & 3.18\%  & 10.33\% & 9.60\%  & 5.46\%  \\
\end{tabular}
\end{ruledtabular}
\end{table*}
\section{Conclusion}\label{sct:conclusion}

This study presents DM-ROM, a novel nonlinear reduced-order modeling (ROM) framework that combines deep learning and manifold learning to predict aerodynamic flow fields accurately. The DM-ROM methodology leverages a CNN-based parameterization network to extract nonlinear shape modes using geometric shape information. The high-dimensional output fields are then reduced using the ISOMAP technique, and an MLP regression model is trained to predict the flow field at unknown design points. The proposed framework is applied to predict the transonic flow field over an RAE2822 airfoil, and the results show that DM-ROM significantly outperforms both POD-ROM and ISOMAP-ROM in terms of prediction accuracy, particularly in capturing highly nonlinear phenomena such as shock waves. DM-ROM notably reduces shock wave location and strength prediction errors, showcasing its capability to handle complex flow interactions more effectively.

An important advantage of DM-ROM is its flexibility. Once the CNN-based parameterization network is trained to extract shape modes, the ISOMAP and MLP regression networks can be easily retrained for different aerodynamic output fields, without the need for repeated shape mode extraction. Although this study has been demonstrated using a 2D airfoil test case, the DM-ROM methodology can be extended to 3D flow problems. Only the CNN-parameterization network would need modification to accommodate 3D shape input, while the ISOMAP and MLP regression components can remain largely unchanged. Furthermore, DM-ROM's approach is not limited to transonic and supersonic aerodynamic flows but can also be applied to other complex flow fields with nonlinear effects, such as shear layer mixing, flow separation, and turbulence modeling. This adaptability makes the framework relevant across a variety of aerodynamic and fluid dynamics problems, as well as other domains involving complex nonlinear interactions.

While DM-ROM exhibits superior performance with larger datasets, one limitation observed in this study is its reduced accuracy when the number of training samples is small. In such cases, linear methods like POD-ROM perform better due to their ability to generalize with fewer data points. To overcome this limitation, future work will focus on developing a multi-fidelity version of DM-ROM, incorporating both high-fidelity and low-fidelity field solutions to improve performance with limited high-fidelity data availability.

\section*{AUTHOR DECLARATIONS}
\subsection*{Conflict of Interest}
The authors have no conflict of interest.


\subsection*{Data Availability Statement}
The data that support the findings of this study are available
from the corresponding author upon reasonable request.

\bibliography{main}
\end{document}